\documentclass[twoside,twocolumn,9pt]{article}
\usepackage{extsizes}
\usepackage[super,sort&compress,comma]{natbib} 
\usepackage[left=1.5cm, right=1.5cm, top=1.785cm, bottom=2.0cm]{geometry}
\usepackage{balance}
\usepackage{mathptmx}
\usepackage{sectsty}
\usepackage{graphicx} 
\usepackage{lastpage}
\usepackage[format=plain,justification=justified,singlelinecheck=false,font={stretch=1.125,small,sf},labelfont=bf,labelsep=space]{caption}
\usepackage{float}
\usepackage{fancyhdr}
\usepackage{fnpos}
\usepackage[english]{babel}
\addto{\captionsenglish}{%
  
}
\usepackage{array}
\usepackage{droidsans}
\usepackage{charter}
\usepackage[T1]{fontenc}
\usepackage[usenames,dvipsnames]{xcolor}
\usepackage{setspace}
\usepackage[compact]{titlesec}
\usepackage{hyperref}

\usepackage{epstopdf}

\definecolor{cream}{RGB}{222,217,201}

\usepackage{mathrsfs,amsmath,amssymb}
\usepackage{threeparttable}
\usepackage{multirow}
\usepackage{booktabs,siunitx}
\usepackage[caption=false]{subfig}

\usepackage[colorinlistoftodos,textsize=normalsize]{todonotes}

\begin{document}

\pagestyle{fancy}
\thispagestyle{plain}
\fancypagestyle{plain}{
\renewcommand{\headrulewidth}{0pt}
}

\makeFNbottom
\makeatletter
\renewcommand\LARGE{\@setfontsize\LARGE{15pt}{17}}
\renewcommand\Large{\@setfontsize\Large{12pt}{14}}
\renewcommand\large{\@setfontsize\large{10pt}{12}}
\renewcommand\footnotesize{\@setfontsize\footnotesize{7pt}{10}}
\makeatother

\renewcommand{\thefootnote}{\fnsymbol{footnote}}
\renewcommand\footnoterule{\vspace*{1pt}%
\color{cream}\hrule width 3.5in height 0.4pt \color{black}\vspace*{5pt}} 
\setcounter{secnumdepth}{5}

\makeatletter 
\renewcommand\@biblabel[1]{#1}            
\renewcommand\@makefntext[1]%
{\noindent\makebox[0pt][r]{\@thefnmark\,}#1}
\makeatother 
\renewcommand{\figurename}{\small{Fig.}~}
\sectionfont{\sffamily\Large}
\subsectionfont{\normalsize}
\subsubsectionfont{\bf}
\setstretch{1.125} 
\setlength{\skip\footins}{0.8cm}
\setlength{\footnotesep}{0.25cm}
\setlength{\jot}{10pt}
\titlespacing*{\section}{0pt}{4pt}{4pt}
\titlespacing*{\subsection}{0pt}{15pt}{1pt}

\fancyfoot{}
\fancyfoot[LO,RE]{\vspace{-7.1pt}\includegraphics[height=9pt]{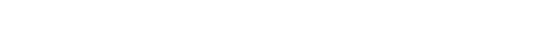}}
\fancyfoot[CO]{\vspace{-7.1pt}\hspace{13.2cm}\includegraphics{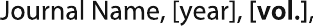}}
\fancyfoot[CE]{\vspace{-7.2pt}\hspace{-14.2cm}\includegraphics{RF}}
\fancyfoot[RO]{\footnotesize{\sffamily{1--\pageref{LastPage} ~\textbar  \hspace{2pt}\thepage}}}
\fancyfoot[LE]{\footnotesize{\sffamily{\thepage~\textbar\hspace{3.45cm} 1--\pageref{LastPage}}}}
\fancyhead{}
\renewcommand{\headrulewidth}{0pt} 
\renewcommand{\footrulewidth}{0pt}
\setlength{\arrayrulewidth}{1pt}
\setlength{\columnsep}{6.5mm}
\setlength\bibsep{1pt}

\makeatletter 
\newlength{\figrulesep} 
\setlength{\figrulesep}{0.5\textfloatsep} 

\newcommand{\topfigrule}{\vspace*{-1pt}%
\noindent{\color{cream}\rule[-\figrulesep]{\columnwidth}{1.5pt}} }

\newcommand{\botfigrule}{\vspace*{-2pt}%
\noindent{\color{cream}\rule[\figrulesep]{\columnwidth}{1.5pt}} }

\newcommand{\dblfigrule}{\vspace*{-1pt}%
\noindent{\color{cream}\rule[-\figrulesep]{\textwidth}{1.5pt}} }

\makeatother

\twocolumn[
  \begin{@twocolumnfalse}
{\includegraphics[height=30pt]{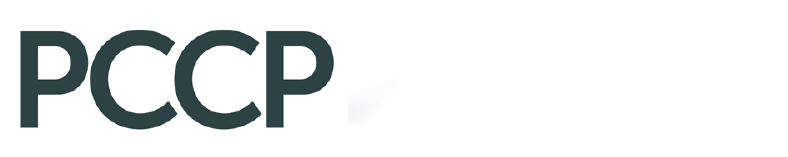}\hfill\raisebox{0pt}[0pt][0pt]{\includegraphics[height=55pt]{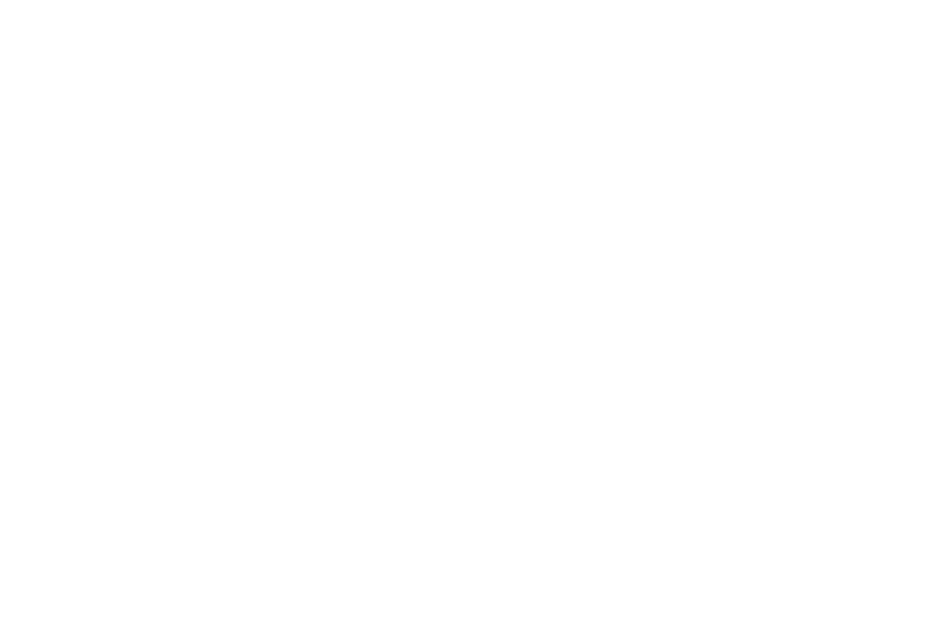}}\\[1ex]
\includegraphics[width=18.5cm]{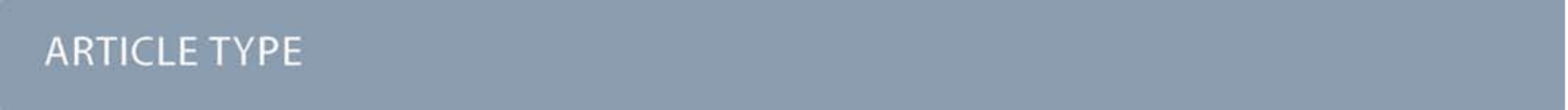}}\par
\vspace{1em}
\sffamily
\begin{tabular}{m{4.5cm} p{13.5cm} }

\includegraphics{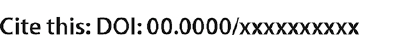} & \noindent\LARGE{\textbf{High-level ab initio quartic force fields and spectroscopic characterization of C$_{2}$N$^{-}$$^\dag$}} \\
\vspace{0.3cm} & \vspace{0.3cm} \\

 & \noindent\large{C.~M.~R.~Rocha$^{\ast}$\textit{$^{a}$} and H.~Linnartz\textit{$^{a}$}} \\

\includegraphics{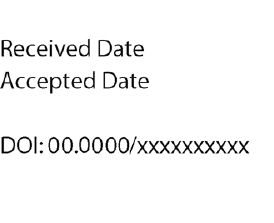} & \noindent\normalsize{While it is now well established that large carbon chain species and radiative electron attachment (REA) are key ingredients triggering 
interstellar anion chemistry, 
the role played 
by smaller 
molecular anions, for which REA 
appears to be an unlikely formation pathway, is as yet elusive. 
Advancing this research 
undoubtedly requires the knowledge (and modeling) of their astronomical 
abundances which, for the case of C$_{2}$N$^{-}$, 
is largely  
hindered by a lack of 
accurate spectroscopic signatures. 
In this work, we provide such 
data for both ground $\ell$-CCN$^{-}$($^{3}\Sigma^{-}$) and low-lying 
$c$-CNC$^{-}$($^{1}A_{1}$) {isomers and their singly-substituted isotopologues} by means of   
state-of-the-art 
rovibrational quantum chemical techniques. 
Their quartic force fields 
are herein calibrated 
using a high-level composite energy 
scheme that accounts for extrapolations to both 
one-particle and (approximate) $\mathcal{N}$-particle basis set limits, 
in addition to relativistic effects, with the final forms being subsequently subject to nuclear motion calculations. Besides standard 
spectroscopic attributes,  
the full set of computed properties includes fine and hyperfine 
interaction constants and can be readily introduced as guesses in 
conventional  
experimental data reduction analyses through effective Hamiltonians. On the basis of benchmark calculations performed anew for a minimal test set of prototypical triatomics and 
limited (low-resolution) experimental data 
for $\ell$-CCN$^{-}$($^{3}\Sigma^{-}$), the 
target 
accuracies 
are determined to be better than 0.1\% of experiment for rotational constants and 0.3\% for vibrational fundamentals. 
Apart from laboratory investigations, the 
results here presented 
are expected to also prompt future astronomical surveys on C$_{2}$N$^{-}$. 
To this end and using the theoretically-{predicted}  
spectroscopic constants, the rotational spectra of both $\ell$-CCN$^{-}$($^{3}\Sigma^{-}$) and $c$-CNC$^{-}$($^{1}A_{1}$) 
are derived and their likely detectability in the interstellar medium 
{\color{black}is} 
further explored in connection with working frequency ranges of powerful astronomical facilities. Our best theoretical estimate places $c$-CNC$^{-}$($^{1}A_{1}$) at about $15.3\,\mathrm{kcal\,mol^{-1}}$ above 
the ground-state $\ell$-CCN$^{-}$($^{3}\Sigma^{-}$) species.  
} \\

\end{tabular}
\end{@twocolumnfalse} \vspace{0.6cm}

  ]

\renewcommand*\rmdefault{bch}\normalfont\upshape
\rmfamily
\section*{}
\vspace{-1cm}


\footnotetext{\textit{$^{a}$~Laboratory for Astrophysics, Leiden Observatory, Leiden University, P.O. Box 9513, NL-2300 RA Leiden, The Netherlands. E-mail: romerorocha@strw.leidenuniv.nl}}

\footnotetext{\dag~Electronic Supplementary Information (ESI) available. See DOI: 00.0000/00000000.}



\section{Introduction}\label{sec:intro}

The plausible existence and role of negative molecular ions in the interstellar medium (ISM) were put forward in the early days of astrochemistry by several authors~\citep{DAL73:95,SAR80:769,HER81:656}. 
However, while their parent cation and neutral species have soon 
emerged as tempting targets for radioastronomical surveys~\citep{MCG018:17} and paved the way for explaining chemical synthesis in the ISM~\citep{SOL72:389,WAT72:321,WAT72:659,HER73:505,DAL76:573}, the detection of anions remained largely elusive, particularly because of a lack of accurate 
rest frequencies~\citep{MOR005:325}.
This situation has changed recently with 
the laboratory and astronomical identification of the first interstellar molecular anion~\citep{MCC06:L141} C$_{6}$H$^{-}$. This led to a resurgence of interest of chemists, physicists, and astrophysicists in anions, motivating new surveys as well as theoretical and laboratory studies~\citep{SIM008:6401,LAR012:066901,FOR015:9941,MIL017:1765}. 
As a result, several other negatively charged
species were soon identified like C$_{4}$H$^{-}$, 
C$_{8}$H$^{-}$, C$_{3}$N$^{-}$, C$_{5}$N$^{-}$ and CN$^{-}$~(
Refs.~\citenum{MIL017:1765},~\citenum{CER020:641} and references therein). 

Very early on, it has been suggested~\citep{SAR80:769,HER81:656} that, under typical interstellar conditions, the 
formation of such anionic inventory ($X^{-}$) 
could be mainly ascribed to radiative electron attachment (REA) to the existing neutrals ($X$)~\citep{HER81:656,BET96:686,PET97:210,TER000:135,MIL000:195,MIL007:L87,HER008:1670,WAL009:752}:
\begin{equation}\label{eq:rea}
X+e^{-}\underset{k_{\text{d}}}{\stackrel{k_{\text{c}}}{\rightleftharpoons}}[X^{-}]^{*}{\stackrel{k_{\text{r}}}{\longrightarrow}} X^{-}+h\nu.     
\end{equation}
As noted elsewhere~\citep{HER81:656,PET97:210,TER000:135,HER008:1670}, reaction~(\ref{eq:rea}) implies a competition, following electron capture ($k_{\text{c}}$), between auto-detachment ($k_{\text{d}}$) and radiative stabilization ($k_{\text{r}}$) of the initially formed (transient)  
superexcited 
complex $[X^{-}]^{*}$; in dilute astrophysical
media, collisional stabilization of $[X^{-}]^{*}$ is assumed negligible~\citep{MIL017:1765}.  
Using phase-space theory (PST) and relying on the mechanism~(\ref{eq:rea}), Herbst~\citep{HER81:656} first derived a theoretical
expression for the overall REA rate constant 
($k_{\text{REA}}$)~--~the results pointed towards an interesting conclusion: $k_{\text{REA}}$ increases greatly with increasing molecular size and electron affinity of the target neutrals ~\citep{HER81:656,PET97:210,TER000:135,HER008:1670}. Moreover, for $X$ species 
with large dipole moments ($\mu\!\gtrsim\!2$-$2.5\,\mathrm{D}$), it has 
later been recognized~\citep{GUT001:466,HER008:1670,CAR013:97,CAR014:054302} that the existence of 
dipole-driven resonances~\citep{CAR013:97,CAR014:054302} and (excited) dipole-bound states~\citep{DES96:1339,SIM008:6401,FOR015:9941} 
are key to enhance further the $[X^{-}]^{*}$'s lifetime (with respect to auto-detachment), and hence the efficiency of REA. 
Thus, provided that the $k_{\text{r}}$'s are high, such 
anions may be formed with sizeable rates ($k_{\text{REA}}\!\sim\!10^{-7}\,\mathrm{cm^{3}\,s^{-1}}$) and, depending on the local
ISM gas pressure and radiation field, exhibit 
appreciable anion-to-neutral ratios~\citep{HER81:656,HER008:1670}. 
{\color{black}It was for this reason that carbon chain anions were also considered possible carriers of diffuse interstellar bands.~\citep{TUL98:L69}}

Assuming reaction~(\ref{eq:rea}) as the major 
anion formation route and using the theoretically-derived PST rates~\citep{HER81:656,PET97:210,TER000:135,HER008:1670}, 
previous 
anion chemical models have been 
successful in reproducing the observed 
abundances of the larger, highly-dipolar carbon-chain anions C$_{n}$H$^{-}$($n\!>\!5$) and C$_{n}$N$^{-}$($n\!>\!4$) in a variety of astronomical environments~\citep{MIL017:1765,MIL000:195,MIL007:L87,WAL009:752}. 
For example,~\citet{WAL009:752} determined anion-to-neutral ratios of~$\sim\!4\%$,~$\sim\!5\%$~and~$\sim\!7\%$~for~C$_{8}$H$^{-}$,~C$_{6}$H$^{-}$~and~C$_{5}$N$^{-}$, respectively, values that compare quite well with the ones observed 
in the dark cloud TMC-1 ($\sim\!5\%$,~$\sim\!2\%$~and~$\sim\!13\%$)~\citep{MIL017:1765,CER020:641}. These results provide further    
evidence in support of 
the REA hypothesis for such molecules. However, for the smallest anionic species  (\emph{e.g.}, CN$^{-}$ and C$_{3}$N$^{-}$) 
for which REA to their parent neutrals are theorized {\color{black}to be} very slow 
($k_{\text{REA}}\!\lesssim\!10^{-10}\,\mathrm{cm^{3}\,s^{-1}}$), notable discrepancies have soon appeared between the modeled and
observed anion-to-neutral ratios~\citep{MIL017:1765}, 
suggesting  
that other alternative pathways might dominate {\color{black}their synthesis~\citep{PET96:137,COR009:68,AGU010:L2,COR012:120,GIA017:42,CER020:641,YUR020:5098}}. 
For example, the unusually  
high CN$^{-}$ abundance observed towards the carbon-rich 
star IRC+10216 has been 
explained~\citep{AGU010:L2} by means of 
the fragmentation reactions C$_{n}^{-}$\!+\!N$\rightarrow$CN$^{-}$\!+\!C$_{n-1}$~\citep{EIC007:1283};   
a similar synthetic route (C$_{n}^{-}$\!+\!N$\rightarrow$C$_{3}$N$^{-}$\!+\!C$_{n-3}$) has been later proposed to also dominate the production of C$_{3}$N$^{-}$ in TMC-1~\citep{CER020:641}. 
{\color{black}Subsequent quantum mechanical calculations 
by~\citet{GIA017:42} provided compelling evidence in favor of the 
H$^{-}$\!+\!HC$_{n}$N reactions as additional prime sources of  
elemental C$_{n}$N$^{-}$'s under typical ISM conditions.} 
Based on laboratory experiments,~\citet{CHA020:90} recently suggested a novel formation pathway for smaller interstellar C$_{n}$N$^{-}$/C$_{n}^{-}$ species~--~it involves the fragmentation decay of 
superexcited resonance anion 
states of larger analogues (\emph{e.g.},  
$[\mathrm{C}_{n}\mathrm{N}^{-}]^{*}$ with $n\!=\!3,5,6,7$) that can be formed  
from impinging UV photons onto the external layers of IRC+10216~[see Eq.~(\ref{eq:rea})]. The results pointed out the dominance of C$_{2}^{-}$ and C$_{2}$N$^{-}$ as 
fragmentation products, thereby offering invaluable
prospects into their omnipresence 
in the circumstellar shells of IRC+10216~\citep{CHA020:90}. Indeed, these species are yet to be identified in space and their laboratory and theoretical characterization is tempting/timely. 
It should be noted that, apart from circumstellar envelopes, the conclusions drawn by~\citet{CHA020:90} are expected to also prompt future 
astronomical surveys on C$_{2}^{-}$ and C$_{2}$N$^{-}$ in strongly shielded environments like TMC-1, although their existence 
therein (if at all) must entail distinct chemical formation routes. 
Thus, studying the astronomical 
abundance of these smaller species is key to a proper understanding of a ISM anion chemistry beyond REA~\citep{AGU010:L2}.
For this, accurate spectral features of such molecules should then be gathered. Still, while C$_{2}^{-}$ is spectroscopically well-characterized in the laboratory~(Ref.~\citenum{NAK017:106} and references therein), 
the amount of theoretical and experimental data on the carbonitrile anion C$_{2}$N$^{-}$ is as yet very 
limited~\citep{PAS99:333,GAR009:064304,FOR017:22860,FRA020:234303}. 
\begin{figure}
{\includegraphics[width=1\linewidth]{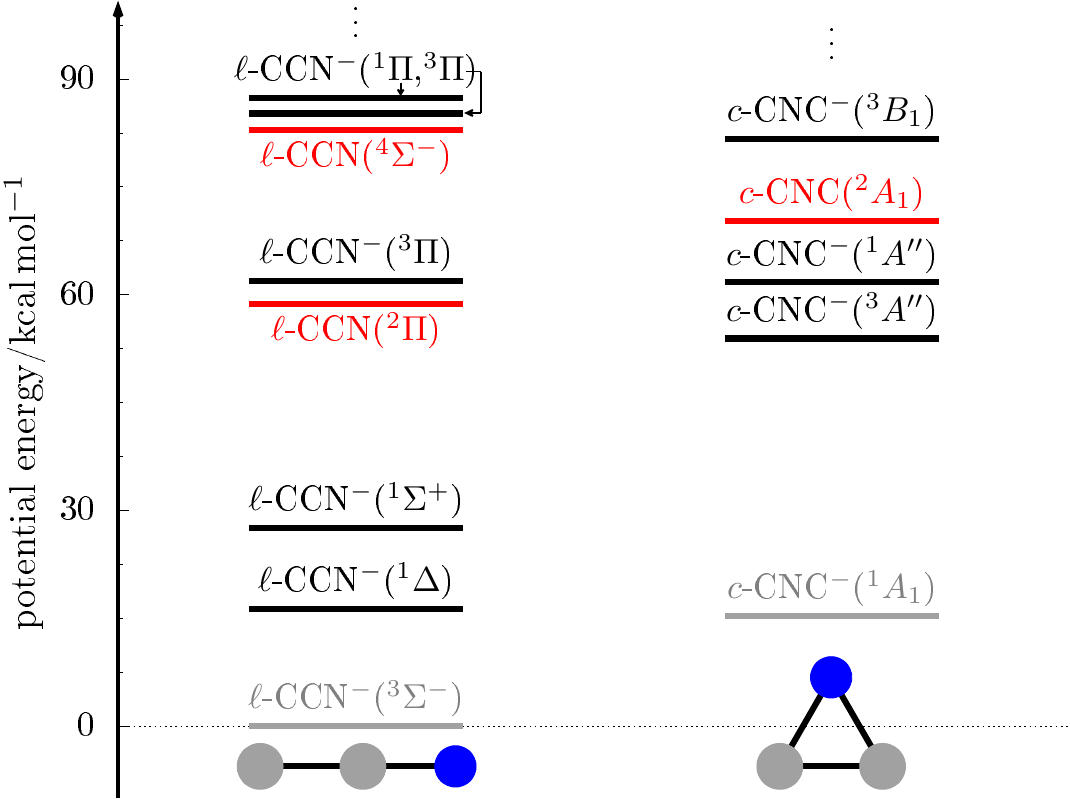}}
\caption{\footnotesize Linear ($\ell$-) and cyclic ($c$-) stationary points on the ground and some low-lying excited potential energy surfaces of C$_{2}$N. Relative energies are obtained at the MRCI(Q)/AV$T$Z//CASSCF/AV$T$Z level of theory. Red and black structures denote the corresponding neutral and anionic excited states, respectively, while the gray lines symbolize the target 
anion forms considered in the present study.
} 
\label{fig:elecstat}
\end{figure}

In light of the foregoing, this work thus aims at providing  
accurate rovibrational spectroscopic constants and anharmonic vibrational
frequencies for C$_{2}$N$^{-}$ by means of a high-level theoretical approach (see below).
Indeed, its parent neutral form, $\ell$-$\mathrm{CCN}(^{2}\Pi)$, has only recently been detected in the circumstellar envelope of IRC+10216~\citep{AND014:L1}. Despite having a relatively small dipole moment ($\mu_{e}\!\approx\!0.3\,\mathrm{D}$~\citep{GRA011:144309}), 
$\ell$-$\mathrm{CCN}(^{2}\Pi)$ is characterized by a 
high (positive) electron affinity, $\mathrm{EA}\!=\!2.7489\pm0.001\,\mathrm{eV}$~\citep{GAR009:064304}, {hence} rendering the corresponding ground-state anion, $\ell$-$\mathrm{CCN}^{-}(^{3}\Sigma^{-})$, exceptionally stable 
with respect to electron loss~\citep{GAR009:064304}; see Figure~\ref{fig:elecstat}. As in the case of C$_{2}^{-}$~\cite{NAK017:106}, this is manifested in the very existence of bound electronically excited C$_{2}$N$^{-}$ states that lie below its photodetachment threshold (Figure~\ref{fig:elecstat}) and that can be optically
connected to $\ell$-$\mathrm{CCN}^{-}(^{3}\Sigma^{-})$. Such a high electron binding energy, in combination with a large dipole moment ($\mu_{e}\!\approx\!2.0\,\mathrm{D}$), makes $\ell$-$\mathrm{CCN}^{-}(^{3}\Sigma^{-})$ amenable 
to observation with new powerful 
instruments such as the atacama large millimeter/submillimeter array (ALMA) and the Green Bank Telescope (GBT). Apart from the linear ground-state, a low-energy cyclic $C_{2v}$ form of C$_{2}$N$^{-}$, $c$-$\mathrm{CNC}^{-}(^{1}A_{1})$, [lying~\emph{ca.}~$15\,\mathrm{kcal\,mol^{-1}}$ above $\ell$-$\mathrm{CCN}^{-}(^{3}\Sigma^{-})$]~\citep{PAS99:333} exists (Figure~\ref{fig:elecstat}) that may {be equally relevant} to interstellar chemistry~\citep{FOR017:22860,MEB002:787} and is likewise 
focus of the present study. Indeed, besides acyclic (small) cyano precursors~\cite{McC021:176,PAR017:452}, there is 
compelling evidence that such elemental N-heterocycles might also play a role 
into the chemical evolution of larger 
astrobiologically-relevant species~\citep{SAN020:4616}. 
In addition to interstellar and 
circumstellar environments, 
we should also mention the likely 
pertinence of these 
nitrile anions to the 
atmosphere of Titan wherein a rich N-based anion chemistry is known to prevail~\citep{VUI009:1558}. 

As for their theoretical spectroscopic characterization, we herein employ the so-called 
quartic force 
field (QFF) approach~\citep{FOR017:22860,SCH005:1106,HUA008:044312,XIN009:104301,FOR015:13,MOR018:1333}. 
Within this framework, the potential energy 
surfaces (PESs)~\citep{ROC019:61} 
of $\ell$-$\mathrm{CCN}^{-}(^{3}\Sigma^{-})$ and $c$-$\mathrm{CNC}^{-}(^{1}A_{1})$ are represented locally by fourth-order Taylor 
series expansions:~\citep{ATT012:273}    
\begin{align}\label{eq:qff}
V(\mathbf{R})=&\frac{1}{2}\sum_{ij}F_{ij}\Delta_{i}\Delta_{j} \nonumber \\
&+\frac{1}{6}\sum_{ijk}F_{ijk}\Delta_{i}\Delta_{j}\Delta_{k} \nonumber \\
&+\frac{1}{24}\sum_{ijkl}F_{ijkl}\Delta_{i}\Delta_{j}\Delta_{k}\Delta_{l},
\end{align}
where $\mathbf{R}\!=\!\{R_{1},R_{2},R_{3}\}$ denotes an arbitrary set of internal coordinates, 
$\Delta_{i}\!=\!R_{i}\!-\!R_{i}^{e}$ {represent coordinate displacements} from the equilibrium geometries $\mathbf{R_{e}}\!=\!\{R_{1}^{e},R_{2}^{e},R_{3}^{e}\}$ and $F_{ij}\!\ldots\!=\!\partial^{n}V/\partial\Delta_{i}\partial\Delta_{j}\!\ldots|_{\Delta_{i,j,\ldots}\!=\!0}$ the force constants; the unrestricted summations run over all possible coordinate indices $3\!\geq\!i,j,k,l\!\geq\!1$. These QFFs will then be {computed} using highly accurate \emph{ab initio} 
energies~\citep{SCH005:1106,HUA008:044312,GRA011:144309}, with the final forms being subsequently subject to nuclear motion calculations~\citep{NIE51:90,MIL72:115,ALI85:1,TEN004:85,SPECTRO}. The details of such a methodology are scrutinized in section~\ref{sec:theor}, while the results are presented in Section~\ref{sec:results}. The astrophysical implications are briefly surveyed in Section~\ref{sec:astro}, with the conclusions being gathered in Section~\ref{sec:conclusions}

\section{Theoretical methods}\label{sec:theor}

\subsection{\emph{Ab initio} calculations \& QFFs}\label{sec:abinitio}
\begin{figure}
\centering
{\includegraphics[width=1\linewidth]{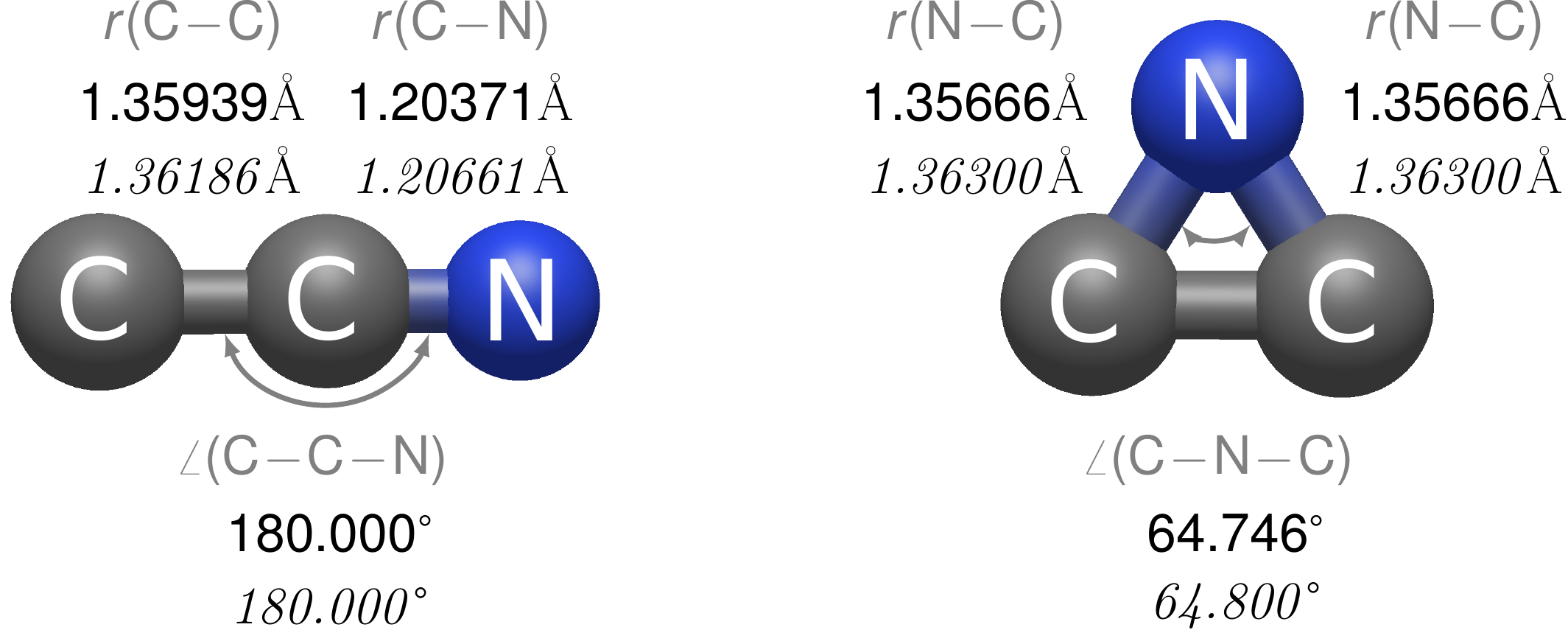}}
\caption{\footnotesize Internal coordinate definitions for~$\ell$-CCN$^{-}$($^{3}\Sigma^{-}$)~and~$c$-CNC$^{-}$($^{1}A_{1}$). Also shown are the corresponding equilibrium (in \textbf{bold}) and vibrationally averaged structures (in \textit{italic}) for the main isotopologues as determined from our final composite QFFs; see later Tables~\ref{tab:rotlin}~and~\ref{tab:rotcyc} for further details.}
\label{fig:struct}
\end{figure}

The full QFFs for the C$_{2}$N anions were computed by performing accurate  
\emph{ab initio} calculations on equally spaced grid points centered 
at best-guess equilibrium structures (see below). A total of 85 symmetry-unique 
geometries were sampled based on a finite (central) difference 
approach; the step lengths taken were $\pm\,0.005\,\si{\angstrom}/\mathrm{rad}$. 
{\color{black}In generating such grids for $\ell$-CCN$^{-}$($^{3}\Sigma^{-}$), we have employed simple internal displacement coordinates~\citep{XIN009:104301}, \emph{i.e.}, 
\begin{align}\label{eq:intcoords}
\Delta_{1}\!=\!r(\mathrm{C_{1}\!-\!N})-r_{e}(\mathrm{C\!-\!N}), \nonumber \\ 
\Delta_{2}\!=\!r(\mathrm{C_{1}\!-\!C_{2}})-r_{e}(\mathrm{C\!-\!C}) 
\end{align}
for stretches and  
\begin{equation}\label{eq:linben}
\Delta_{3,4}\!=\!\sin\left[\angle(\mathrm{C\!-\!C\!-\!N})\right]-\sin\left[\angle_{e}(\mathrm{C\!-\!C\!-\!N})\right] 
\end{equation}
for the degenerate linear bends; $r$ and $\angle$ define bond lengths and angle, respectively, with the subscript $e$ denoting the corresponding equilibrium values; see Figure~\ref{fig:struct}. Note that only one component, $\Delta_{3}$, {\color{black}was} considered
in the finite difference calculations; bending force constants 
depending on $\Delta_{4}$ 
are herein determined 
via cylindrical symmetry 
relations~\citep{HOY72:1265}   
\begin{align}\label{eq:cilindrical}
F_{3344}&=(F_{333}+4F_{33})/3, & F_{33ij}&=F_{44ij}, &  \nonumber \\ 
F_{33i}&=F_{44i}, & F_{33}&=F_{44} & \forall i,j. 
\end{align}
For $c$-CNC$^{-}$($^{1}A_{1}$), the following 
symmetry-internal displacement coordinates were used~\citep{FOR015:13}:
\begin{align}\label{eq:symcoords}
\Delta_{1}&=\frac{1}{\sqrt{2}}[r(\mathrm{N\!-\!C_{1}})+r(\mathrm{N\!-\!C_{2}}) -2r_{e}(\mathrm{N\!-\!C})] \nonumber \\
\Delta_{2}&=\angle(\mathrm{C\!-\!N\!-\!C})-\angle_{e}(\mathrm{C\!-\!N\!-\!C}) \nonumber \\
\Delta_{3}&=\frac{1}{\sqrt{2}}[r(\mathrm{N\!-\!C_{1}})-r(\mathrm{N\!-\!C_{2}})].
\end{align}
}
At each selected geometry $\mathbf{R}$, the total electronic energy, $E$, was then obtained via a composite scheme~\citep{SCH005:1106,HUA008:044312,GRA011:144309}  
\begin{equation}\label{eq:energy}
E(\mathbf{R})=E^{\mathrm{CC}}_{\infty}(\mathbf{R})+\Delta_{\mathrm{DKH}}(\mathbf{R})+\Delta_{\mathrm{HO}}(\mathbf{R}), 
\end{equation}
where $E^{\mathrm{CC}}_{\infty}$ is an estimate of the one-particle complete basis set (CBS) limit~\citep{VAR018:177}, including core and core-valence correlation, at the coupled cluster singles and doubles level of theory with perturbative triples~\citep{RAG89:479} [CCSD(T) or, briefly, CC], $\Delta_{\mathrm{DKH}}$ is a correction for scalar relativistic effects~\citep{PEN012:1081}, and $\Delta_{\mathrm{HO}}$ accounts for 
higher-order (HO) electron correlation contributions beyond CC. 
All calculations have been performed at the spin-restricted (open-shell) CC level of theory~\citep{RAG89:479,KNO93:5219,WAT93:8718} using the restricted (open-shell) Hartree-Fock (HF) determinant as reference. 
The V$X$Z ($X\!=\!D,T,Q,5$) basis sets of Dunning and co-workers~\citep{DUN89:1007} with additional 
diffuse~\citep{KEN92:6796} (AV$X$Z) and core correlation  
functions~\citep{WOO95:4572} (ACV$X$Z) were employed throughout, 
with the computations done with \texttt{MOLPRO}~\citep{MOLPRO}. 
To ensure accuracy of the final force constants, all calculations have been carried out with a  
convergence energy criteria of $10^{-12}\,\si{\hartree}$~\cite{SCH005:1106,HUA008:044312}.  

Due to the distinct asymptotic 
convergence rates~\citep{VAR018:177}, the CBS extrapolations for the HF and total CC electron correlation (cor) components of $E^{\mathrm{CC}}_{\infty}$ [Eq.~(\ref{eq:energy})] were performed individually, \emph{i.e.}, 
\begin{equation}\label{eq:cbsenergy}
E^{\mathrm{CC}}_{\infty}(\mathbf{R})=E^{\mathrm{HF}}_{\infty}(\mathbf{R})+E^{\mathrm{cor}}_{\infty}(\mathbf{R}). 
\end{equation}
For the HF energy, a three-point exponential-type formula has been so employed~\citep{FEL93:7059} 
\begin{equation}\label{eq:cbshf}
E^{\mathrm{HF}}_{X}(\mathbf{R})=E^{\mathrm{HF}}_{\infty}(\mathbf{R})+A\exp{(-BX)}, 
\end{equation}
where $E^{\mathrm{HF}}_{\infty}$, $A$, 
and $B$ are parameters to be calibrated from HF/ACV$X$Z ($X\!=\!T,Q,5$) energies. In turn, the extrapolated cor contributions  
are obtained via the inverse-power formula~\citep{VAR007:244105}   
\begin{equation}\label{eq:cbscor}
E^{\mathrm{cor}}_{X}(\mathbf{R})=E^{\mathrm{cor}}_{\infty}(\mathbf{R})+ A'X^{-3}+B'X^{-5}, 
\end{equation}
where $E^{\mathrm{cor}}_{\infty}$, $A'$ and $B'$ are calibrated 
from the raw CC/ACV$X$Z ($X\!=\!T,Q,5$) total correlation energies. 

In Eq.~(\ref{eq:energy}), the corrections due to scalar relativistic 
contributions were obtained by
\begin{equation}\label{eq:rel}
\Delta_{\mathrm{DKH}}(\mathbf{R})=E_{\mathrm{DKH}}(\mathbf{R})-E_{\mathrm{NR}}(\mathbf{R}), 
\end{equation}
where $E_{\mathrm{DKH}}$ is the total second-order Douglas-Kroll-Hess (DKH)~\citep{DOU74:89,JAN89:6016,PEN012:1081}
CC 
energy calculated with the V$T$Z-DK basis set~\citep{JON001:48}; $E_{\mathrm{NR}}$ defines its non-relativistic CC/V$T$Z-DK counterpart. As for the estimation of higher-order electron correlation, $\Delta_{\mathrm{HO}}$ [Eq.~(\ref{eq:energy})], we herein include three additional (frozen-core) energy increments~\citep{FEL006:054107,FEL008:204105,GRA011:144309}
\begin{equation}\label{eq:ho}
\Delta_{\mathrm{HO}}(\mathbf{R})=\Delta\mathrm{T}(\mathbf{R})+\Delta\mathrm{Q}(\mathbf{R})+\Delta\mathrm{FCI}(\mathbf{R}),  
\end{equation}
where $\Delta\mathrm{T}$ and $\Delta\mathrm{Q}$ account for the residual   
CC correlation components associated with iterative triple ($T_{3}$) and 
quadruple ($T_{4}$) excitations; the first is obtained via differences in energy between CCSDT~\citep{NOG87:7041,SCU88:382} and CCSD(T) calculations with the V$Q$Z basis set, while the second is determined from CCSDTQ~\citep{KUC92:4282,OLI91:1229}/V$D$Z and CCSDT/V$D$Z energy differences. The importance 
of such corrections (and cost-effective variants) for accurate predictions of spectroscopic/thermochemical properties of small-to-medium sized molecules has been emphasized in several previous works~\citep{TAJ004:11599,SCH005:1106,FEL008:204105,KAR010:144102,MOR018:1333,MOR018:3483,KAR020:024102,PUZ020:6507}. Furthermore, to account for the small,
residual errors arising from the truncation of the $\mathcal{N}$-particle 
expansions at the CCSDTQ level, we also include in Eq.~(\ref{eq:ho}) an estimate of the difference in correlation energy between CCSDTQ and full configuration interaction (FCI), $\Delta\mathrm{FCI}$, calculated with the V$D$Z basis set ($X\!=\!D$). The FCI limit was then obtained via a continued
fraction (cf) approximant~\citep{GOO002:6948,FEL006:054107} 
\begin{equation}\label{eq:cf}
E^{\mathrm{FCI}}_{X}(\mathbf{R})\approx\frac{E^{\mathrm{CCSD}}_{X}}{1-\left[\left(\frac{\delta_{\mathrm{T}}}{E^{\mathrm{CCSD}}_{X}}\right)\Big/\left(1-\frac{\delta_{\mathrm{Q}}}{\delta_{\mathrm{T}}}\right) \right]}, 
\end{equation}
where $\delta_{\mathrm{T}}\!=\!E^{\mathrm{CCSDT}}_{X}\!-\!E^{\mathrm{CCSD}}_{X}$ and $\delta_{\mathrm{Q}}\!=\!E^{\mathrm{CCSDTQ}}_{X}\!-\!E^{\mathrm{CCSDT}}_{X}$. Thus, $\Delta\mathrm{FCI}$ in Eq.~(\ref{eq:ho}) is defined as $E^{\mathrm{FCI}}_{X}-E^{\mathrm{CCSDTQ}}_{X}$. 
For systems with small-to-moderate multireference character [the $\mathcal{T}_{1}$ diagnostic values for $\ell$-$\mathrm{CCN}^{-}(^{3}\Sigma^{-})$ and $c$-$\mathrm{CNC}^{-}(^{1}A_{1})$ 
are $\approx\!0.025$~and~$0.012$, respectively], Eq.~(\ref{eq:cf}) has shown to be a viable alternative 
for estimating electron correlation beyond CCSDTQ, recovering nearly 80\% of available FCI corrections~\citep{FEL008:204105,FEL006:054107}. 
It should be noted, however, that, while
the use of the CCSDTQP method~\citep{KAL005:214105} and larger basis sets, \emph{e.g.}, V$T$Z, would be preferable in estimating 
$E^{\mathrm{FCI}}_{X}$ (see, \emph{e.g.}, Ref.~\citenum{FEL008:204105}) and $\Delta\mathrm{Q}$, respectively, the associated computational cost would make the task of calculating the QFFs intractable with current available resources; all HO corrections have been computed with the \texttt{MRCC}~\citep{MRCC} code.

\begin{table}[htb!]
\centering
\caption{\footnotesize Internal coordinate force constants for the C$_{2}$N anions as determined from our final composite QFFs [Eqs.~(\ref{eq:qff})~and~(\ref{eq:energy})]. Units are $\mathrm{mdyn}\,\si{\angstrom}^{-n}\si{\radian}^{-m}$ appropriate for an energy unit of $\mathrm{mdyn}\,\si{\angstrom}(\equiv\,\si{\atto\joule})$. 
Eqs.~(\ref{eq:intcoords})-(\ref{eq:symcoords}) define the coordinates.
}
\label{tab:force}
\begin{threeparttable}
\begin{tabular}{
l
S[table-align-text-post=false,table-format=4.6]
c
S[table-align-text-post=false,table-format=4.6]
S[table-align-text-post=false,table-format=4.6]}
\hline\hline \\[-1.8ex]
  & {$\ell$-CCN$^{-}$($^{3}\Sigma^{-}$)} & & \multicolumn{2}{c}{$c$-CNC$^{-}$($^{1}A_{1}$)}                               \\
  \cline{2-2} \cline{4-5}\\[-1.8ex]
  & {QFF\tnote{a}} & & {QFF\tnote{a}} & {QFF\tnote{b}} \\[0.5ex] 
\hline \\[-2.25ex]
$F_{11}$                  &  12.364302 & &   8.601837 &   8.613181 \\
$F_{21}$                  &   2.083929 & &   3.701487 &   3.697293 \\
$F_{22}$                  &   5.491370 & &   6.094652 &   6.098824 \\
$F_{33/44\tnote{c}}$      &   0.414145 & &   5.415849 &   5.451568 \\
$F_{111}$                 & -94.3113   & & -37.2385   & -37.1905   \\
$F_{211}$                 &   2.4761   & & -11.6437   & -11.5930   \\
$F_{221}$                 &  -6.9087   & & -23.3909   & -23.3002   \\
$F_{331/441\tnote{c}}$    &  -0.8749   & & -26.7040   & -26.6360   \\
$F_{222}$                 & -38.6440   & & -46.5351   & -46.2960   \\
$F_{332/442\tnote{c}}$    &  -0.4844   & &  -2.0261   &  -1.9450   \\  
$F_{1111}$                & 497.67     & & 133.99     & 133.55     \\
$F_{2111}$                &  21.11     & &  29.28     &  29.29     \\
$F_{2211}$                & -17.79     & &  54.40     &  54.25     \\
$F_{3311/4411\tnote{c}}$  &  -0.30     & & 103.14     & 102.91     \\
$F_{2221}$                &  16.36     & & 119.50     & 118.99     \\
$F_{3321/4421\tnote{c}}$  &   2.43     & & -18.96     & -18.22     \\
$F_{2222}$                & 207.87     & & 316.24     & 313.96     \\
$F_{3322/4422\tnote{c}}$  &  -0.75     & & -41.59     & -40.55     \\
$F_{3333/4444\tnote{c}}$  &   2.16     & &  74.07     &  73.89     \\
$F_{3344}\tnote{c}$       &   1.27     & &            &            \\
\hline\hline
\end{tabular}
\begin{tablenotes}[flushleft]
  \item[a]{{\footnotesize This work}.}   
  \item[b]{{\footnotesize Ref.~\citenum{FOR017:22860}}.}
  \item[c]{{\footnotesize Only relevant for $\ell$-CCN$^{-}$; see Eq.~(\ref{eq:cilindrical})}.}
\end{tablenotes}
\end{threeparttable}
\end{table}
Due to a lack of accurate experimental geometries for C$_{2}$N$^{-}$, the determination of the
reference structures in which to expand our QFFs relied solely on high-level \emph{ab initio} estimates~\citep{HUA008:044312}. This has been done 
by first optimizing geometries at the CC/ACV$T$Z level, followed by computations of cost-effective QFFs therein using CBS-extrapolated CC/ACV$X$Z ($X\!=\!D,T,Q$) plus $\Delta_{\mathrm{DKH}}$ energies [Eqs.~(\ref{eq:cbsenergy})-(\ref{eq:rel})]. The resulting fine-tuned minima were then utilized as reference for final geometry displacements and 
energy evaluations via Eq.~(\ref{eq:energy}). Of course, as these optimum configurations are not exact minima on the final composite PESs, accurate QFFs and equilibrium geometries could only be obtained by least-squares fitting such a composite energy set to Eq.~(\ref{eq:qff}); the sum of squared residuals were typically $10^{-15}\,\si{\hartree}^{2}$, with the resulting force constants being  numerically defined in Table~\ref{tab:force}.
\begin{figure*}[htb!]
\centering
{\includegraphics[width=0.8\linewidth]{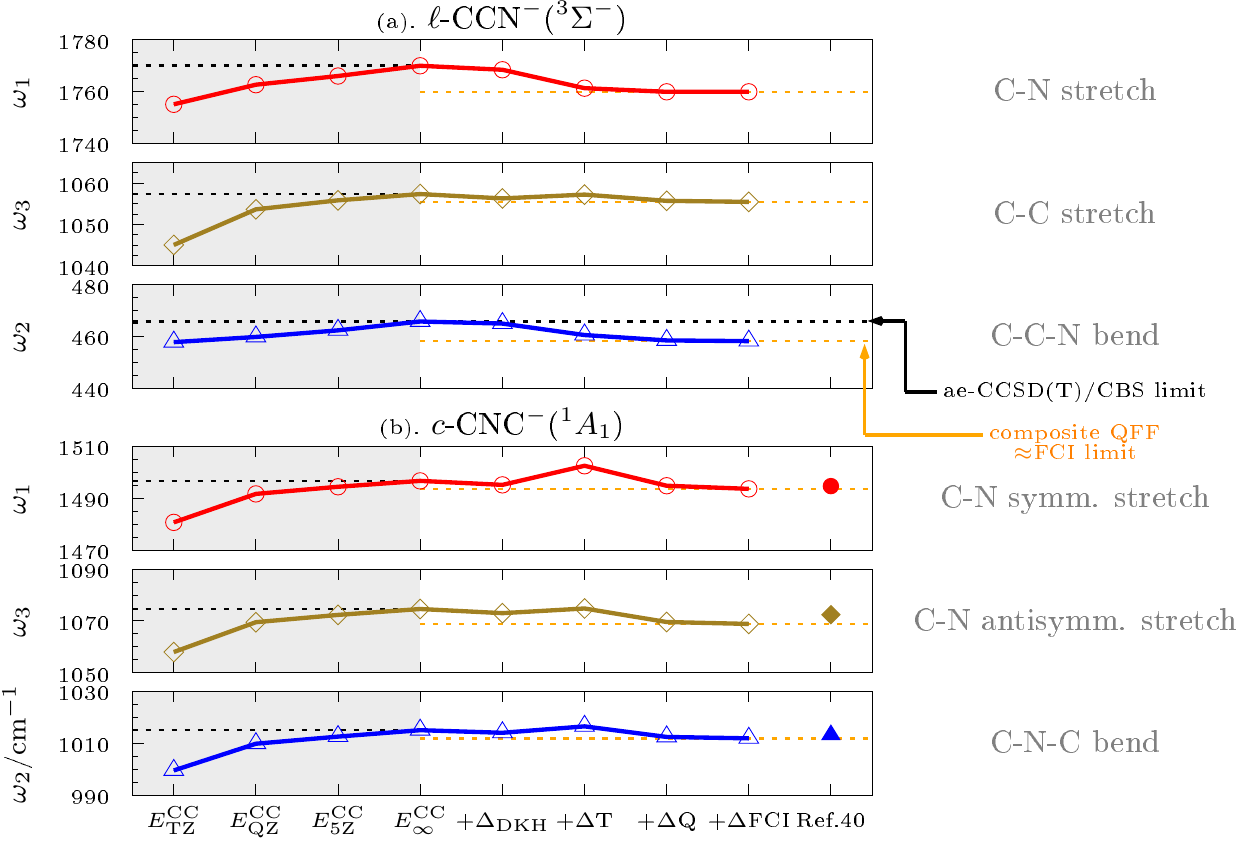}}
\caption{\footnotesize Convergence of the predicted harmonic frequencies (in cm$^{-1}$) for~(a).~$\ell$-CCN$^{-}$($^{3}\Sigma^{-}$)~and~(b).~$c$-CNC$^{-}$($^{1}A_{1}$) as a function of each energy increment in Eq.~(\ref{eq:energy}). For comparison, we also show the calculated values from the raw CC/ACV$X$Z ($X\!=\!T,Q,5$) QFFs separately as well as the most accurate results from the literature~\citep{FOR017:22860} for the $c$-CNC$^{-}$ species [panel~(b)]. Shaded gray areas mark the transition 
region from one-particle to $\mathcal{N}$-particle expansion extrapolations. Black and orange lines outline the corresponding frequencies obtained at the one-particle CBS limit ($E^{\mathrm{CC}}_{\infty}$) and from the final QFFs, respectively; ae stands for all electron (non-frozen-core) values.}
\label{fig:convergence}
\end{figure*}

\subsection{(Ro)Vibrational calculations and spectroscopic constants}\label{sec:vib}
With such QFFs at hand, the correspoding rovibrational energy levels and associated spectroscopic constants can then be determined by solving the 
nuclear Schr\"{o}dinger equation (NSE)~\citep{ROC019:61}. This has been here accomplished through standard second-order perturbation theory (VPT2)~\citep{NIE51:90,MIL72:115,ALI85:1} as implemented in \texttt{SPECTRO}~\citep{SPECTRO}. As usual~\citep{FOR015:13,FOR017:22860}, \texttt{SPECTRO} requires the input of the appropriate resonances; for $\ell$-CCN$^{-}$($^{3}\Sigma^{-}$), they correspond to Fermi type-1 ($2\nu_{2}\!\approx\!\nu_{3}$), 
while type-C Coriolis ($\nu_{2}\!\approx\! \nu_{3}$) and Darling-Dennison ($2\nu_{2}\!\approx\!2\nu_{3}$) {\color{black}resonances are} input for $c$-CNC$^{-}$($^{1}A_{1}$). Besides VPT2, rovibrational band origins were also obtained using the exact kinetic energy nuclear motion code \texttt{DVR3D}~\citep{TEN004:85} which computes variationally exact solutions to the three-atom NSE within the framework of the discrete variable representation~\citep{TEN004:85}; {sample \texttt{SPECTRO} inputs 
and the parameters employed in \texttt{DVR3D} are given in the Electronic Supplementary Information~(ESI).} 
Note that, to avoid non-physical results 
and ensure the correct limiting behavior of the PESs, 
the QFFs of $\ell$-CCN$^{-}$($^{3}\Sigma^{-}$) and $c$-CNC$^{-}$($^{1}A_{1}$) 
have been analytically transformed into Morse-sine and Morse-cosine coordinate representations, respectively, prior to the variational \texttt{DVR3D} calculations (VAR); the reader is addressed to Refs.~\citenum{DAT94:5853}~and~\citenum{FOR013:1} for further details.   

\subsection{Benchmark calculations}\label{sec:bench}
To assess the performance of the above protocol, preliminary 
benchmark calculations have been carried out for a limited  
set of triatomics for which accurate gas-phase experimental data 
are available. The selected targets 
comprise the prototypical $\ell$-$\mathrm{HCN}(^{1}\Sigma^{+})$ and $c$-$\mathrm{H_{2}O}(^{1}A_{1})$ molecules as well as $\ell$-$\mathrm{CCO}(^{3}\Sigma^{-})$, an open-shell species that 
is isoelectronic to C$_{2}$N$^{-}$, hence expected to show a similar electronic structure; the final force constants and detailed data analysis are presented in Tables~S2-S4. 
The results indicate that our present  
methodology is capable of producing vibrationally-averaged rotational constants 
and vibrational fundamentals to within $\sim\!0.1\%\,(22\,\mathrm{MHz})$ and $\sim\!0.3\%\,(3\,\mathrm{cm^{-1}})$ of experiment, respectively, for species with at least two heavy atoms, 
{hence} further showcasing its reliability. This is about the accuracy 
one might expect 
for the predicted spectroscopic attributes of $\ell$-CCN$^{-}$($^{3}\Sigma^{-}$) and $c$-CNC$^{-}$($^{1}A_{1}$) 
and is quite consistent with well-established state-of-the-art QFF/VPT2 protocols  
currently available in the literature~\citep{SCH005:1106,MOR018:1333,PUZ020:6507,GAR021:119184}. 

\section{Results}\label{sec:results}
Figure~\ref{fig:convergence} displays the dependence    
of the $\ell$-CCN$^{-}$($^{3}\Sigma^{-}$) and $c$-CNC$^{-}$($^{1}A_{1}$) harmonic frequencies {\color{black}for the three fundamental modes} ($\omega_{i}$) on the ACV$X$Z basis set size at the CC level as well as upon inclusion of the {\color{black}$\Delta_{\mathrm{DKH}}$ [Eq.~(\ref{eq:rel})] and $\Delta_{\mathrm{HO}}$ [Eq.~(\ref{eq:ho})] energy increments [Eq.~(\ref{eq:energy})]}; the corresponding profiles obtained for 
equilibrium geometries ($R_{i}^{e}$) are depicted in Figure~S1 {\color{black}(see also Figure~\ref{fig:struct} to assess their final values)}. Tables~\ref{tab:rotlin}-\ref{tab:anharm} gather the calculated rovibrational spectroscopic constants, vibrational fundamentals and anharmonic constants 
for the various 
C$_{2}$N$^{-}$ forms as obtained from our final 
composite QFFs and VPT2/VAR. {\color{black}Note that, apart from the main isotopologues, detailed spectroscopic data are also presented 
for the $^{13}$C and $^{15}$N singly-substituted species.}

{\color{black}\subsection{Effects of various corrections on equilibrium properties}\label{conv}}
\begin{table*}[htb!]
\centering
\caption{\footnotesize Equilibrium structures and spectroscopic vibration-rotation constants of $\ell$-CCN$^{-}$($^{3}\Sigma^{-}$) isotopologues. Data determined from our final composite QFF (Table~\ref{tab:force}) via second-order perturbation theory (VPT2)~\citep{NIE51:90,MIL72:115,ALI85:1}. 
The bottom part gathers the \emph{ab initio} calculated fine and leading 
hyperfine interaction constants at the QFF equilibrium geometry. Units are MHz unless stated otherwise.}
\label{tab:rotlin}                                                                                                                                                                                                                                                                                                                  
\begin{threeparttable}
\begin{tabular}{
l
S[table-align-text-post=false,table-format=5.4]
c
S[table-align-text-post=false,table-format=5.4]
c
S[table-align-text-post=false,table-format=5.4]
c
S[table-align-text-post=false,table-format=5.4]}
\hline\hline \\[-1.8ex]
  {\multirow[c]{2}{*}{\includegraphics[width=0.09\textwidth]{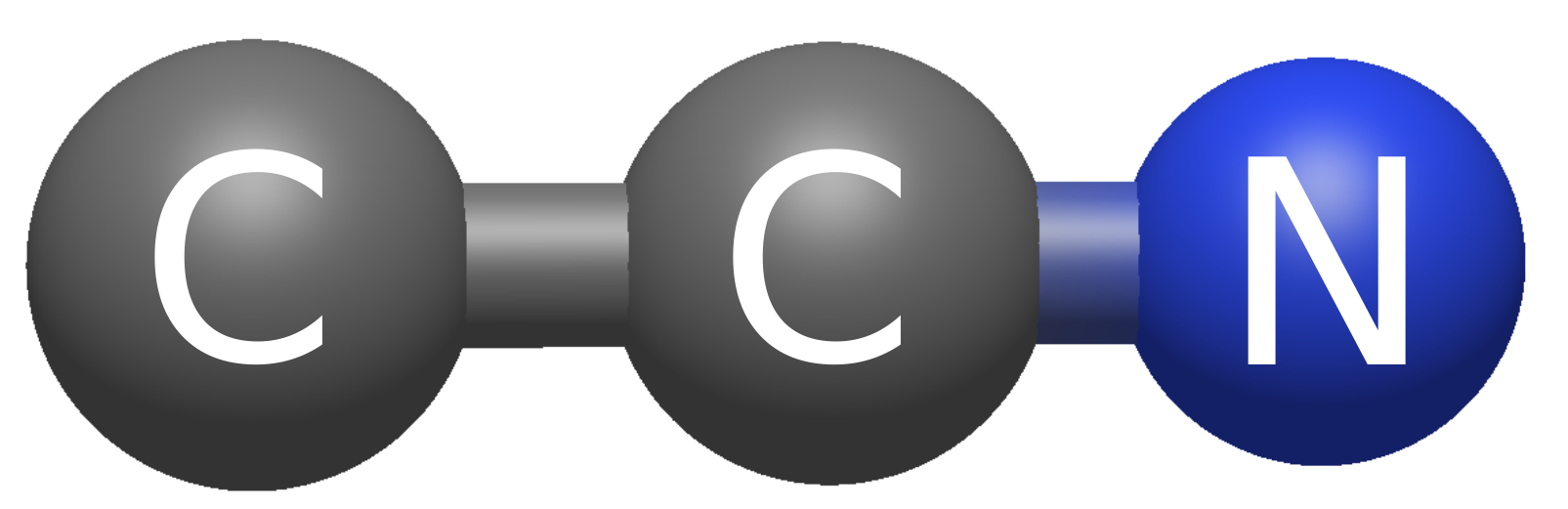}}} & \multicolumn{1}{c}{$\ell$-CCN$^{-}$} & & \multicolumn{1}{c}{$\ell$-$^{13}$CCN$^{-}$} & & \multicolumn{1}{c}{$\ell$-C$^{13}$CN$^{-}$} & & \multicolumn{1}{c}{$\ell$-CC$^{15}$N$^{-}$} \\
\cline{2-2}\cline{4-4}\cline{6-6}\cline{8-8}\\[-2.25ex]
  & {QFF\tnote{a}} & & {QFF\tnote{a}} & & {QFF\tnote{a}} & & {QFF\tnote{a}} \\[0.5ex] 
\hline \\[-2.25ex]
$r_{0}(\mathrm{C\!-\!C})$/\si{\pico\metre}        &   136.1860 & &   136.1765       & &   136.1882       & &   136.1839 \\
$r_{0}(\mathrm{C\!-\!N})$/\si{\pico\metre}        &   120.6613 & &   120.6565       & &   120.6562       & &   120.6600 \\
$\angle_{0}(\mathrm{C\!-\!C\!-\!N})$/\si{\degree} &   180.0000 & &   180.0000       & &   180.0000       & &   180.0000 \\
$B_{0}$                                           & 11861.91   & & 11369.30         & & 11862.02         & & 11490.06   \\
$B_{1}$                                           & 11796.21   & & 11307.99         & & 11800.31         & & 11423.77   \\
$B_{2}$                                           & 11893.37   & & 11399.60         & & 11891.87         & & 11520.70   \\
$B_{3}$                                           & 11795.79   & & 11309.84         & & 11791.59         & & 11429.12   \\
$10^{3}{D}_{e}$                                   &     6.130  & &     5.692        & &     6.128        & &     5.700  \\
$10^{6}{H}_{e}$                                   &    -0.002  & &    -0.002        & &    -0.002        & &    -0.002  \\
${q}$                                             &   -28.489  & &   -26.346        & &   -29.257        & &   -26.867  \\
$\alpha^{B}_{1}$                                  &    65.7    & &    61.3          & &    61.7          & &    66.3    \\
$\alpha^{B}_{2}$                                  &   -31.5    & &   -30.3          & &   -29.8          & &   -30.6    \\
$\alpha^{B}_{3}$                                  &    80.2    & &    76.3          & &    80.7          & &    75.2    \\
$r_{e}(\mathrm{C\!-\!C})$/\si{\pico\metre}        &   135.9387 & &     {-}          & &     {-}          & &     {-}    \\
$r_{e}(\mathrm{C\!-\!N})$/\si{\pico\metre}        &   120.3706 & &     {-}          & &     {-}          & &     {-}    \\
$\angle_{e}(\mathrm{C\!-\!C\!-\!N})$/\si{\degree} &   180.0000 & &     {-}          & &     {-}          & &     {-}    \\
$B_{e}$                                           & 11903.44   & & 11407.80         & & 11903.38         & & 11530.17   \\
$\mu_{e}$/$D$\tnote{b}                            &     1.9873 & &     1.8190       & &     1.9869       & &     2.1347 \\
\hline\\[-2.25ex]
$\lambda_{e}$\tnote{c}                            & 11958.87   & &     {-}          & &     {-}          & &     {-}    \\
$\gamma_{e}$\tnote{c}                             &   -19.16   & &   -18.37         & &   -19.16         & &   -18.56   \\
$b_{F}(\mathrm{^{14}N})$\tnote{d}                 &    19.84   & & {$19.84(65.03)$} & & {$19.84(-49.72)$}& &     {-}    \\
$c(\mathrm{^{14}N})$\tnote{d}                     &   -16.24   & &{$-16.24(-56.62)$}& &{$-16.24(15.80)$} & &     {-}    \\
$eQq(\mathrm{^{14}N})$\tnote{e}                   &    -2.9088 & &     {-}          & &     {-}          & &            \\
$\eta(\mathrm{^{14}N})$\tnote{e}                  &     0.0000 & &     {-}          & &     {-}          & &            \\
\hline\hline
\end{tabular}
\begin{tablenotes}[flushleft]
  \item[a]{{\footnotesize This work. Data obtained using \texttt{SPECTRO}~\citep{SPECTRO}. Rotational constants ($B_{i}$), vibration-rotation interaction constants ($\alpha^{B}_{i}$), quartic (${D}_{e}$) and sextic (${H}_{e}$) centrifugal distortion parameters and $\ell$-type doubling constant ($q$) are all in MHz. {$B_{i}$ with $i\!=\!1$-$3$ are  effective rotational constants calculated for the three vibrational fundamentals; see Figure~\ref{fig:convergence} for mode descriptions. The corresponding zero-point level constant is $B_{0}$}}.}
  \item[b]{{\footnotesize CBS-extrapolated 
  dipole moments (in D) at the QFF equilibrium geometry; see Ref.~\citenum{CON020:024105}}.}
  \item[c]{{\footnotesize Spin-spin ($\lambda_{e}$) and spin-rotation ($\gamma_{e}$) coupling constants (in MHz); see text}.}
  \item[d]{{\footnotesize Electron spin-nuclear spin hyperfine coupling constants. 
  Isotropic Fermi-contact ($b_{F}$) and anisotropic dipole-dipole ($c$) magnetic couplings (both in MHz) evaluated at the $\mathrm{^{14}N}$ nucleus. The corresponding values obtained at $\mathrm{^{13}C}$ are given in parenthesis; couplings at $\mathrm{^{15}N}$ are not explicitly considered}.}  
  \item[e]{{\footnotesize CBS-extrapolated 
  $\mathrm{^{14}N}$ nuclear quadrupole (hyperfine) coupling constant ($eQq$ in MHz) and asymmetry parameter ($\eta$  unitless); see text}.} 
\end{tablenotes}
\end{threeparttable}
\end{table*}
\begin{table*}[htb!]
\centering
\caption{\footnotesize Equilibrium structures, A-reduced Hamiltonian spectroscopic constants (in the I$^{r}$ representation), and
vibration-rotation interaction constants of $c$-CNC$^{-}$($^{1}A_{1}$) isotopologues. Data determined from our final composite QFF (Table~\ref{tab:force}) via second-order perturbation theory (VPT2)~\citep{NIE51:90,MIL72:115,ALI85:1}. The bottom part gathers the \emph{ab initio} calculated leading hyperfine interaction constants at the QFF equilibrium geometry. Units are MHz unless stated otherwise.}
\label{tab:rotcyc}
\begin{threeparttable}
\begin{tabular}{
l
S[table-align-text-post=false,table-format=5.4]
S[table-align-text-post=false,table-format=5.4]
c
S[table-align-text-post=false,table-format=5.4]
S[table-align-text-post=false,table-format=5.4]
c
S[table-align-text-post=false,table-format=5.4]
S[table-align-text-post=false,table-format=5.4]}
\hline\hline \\[-1.8ex]
 {\multirow[b]{2}{*}{\hspace*{0.4cm}\includegraphics[width=0.065\textwidth]{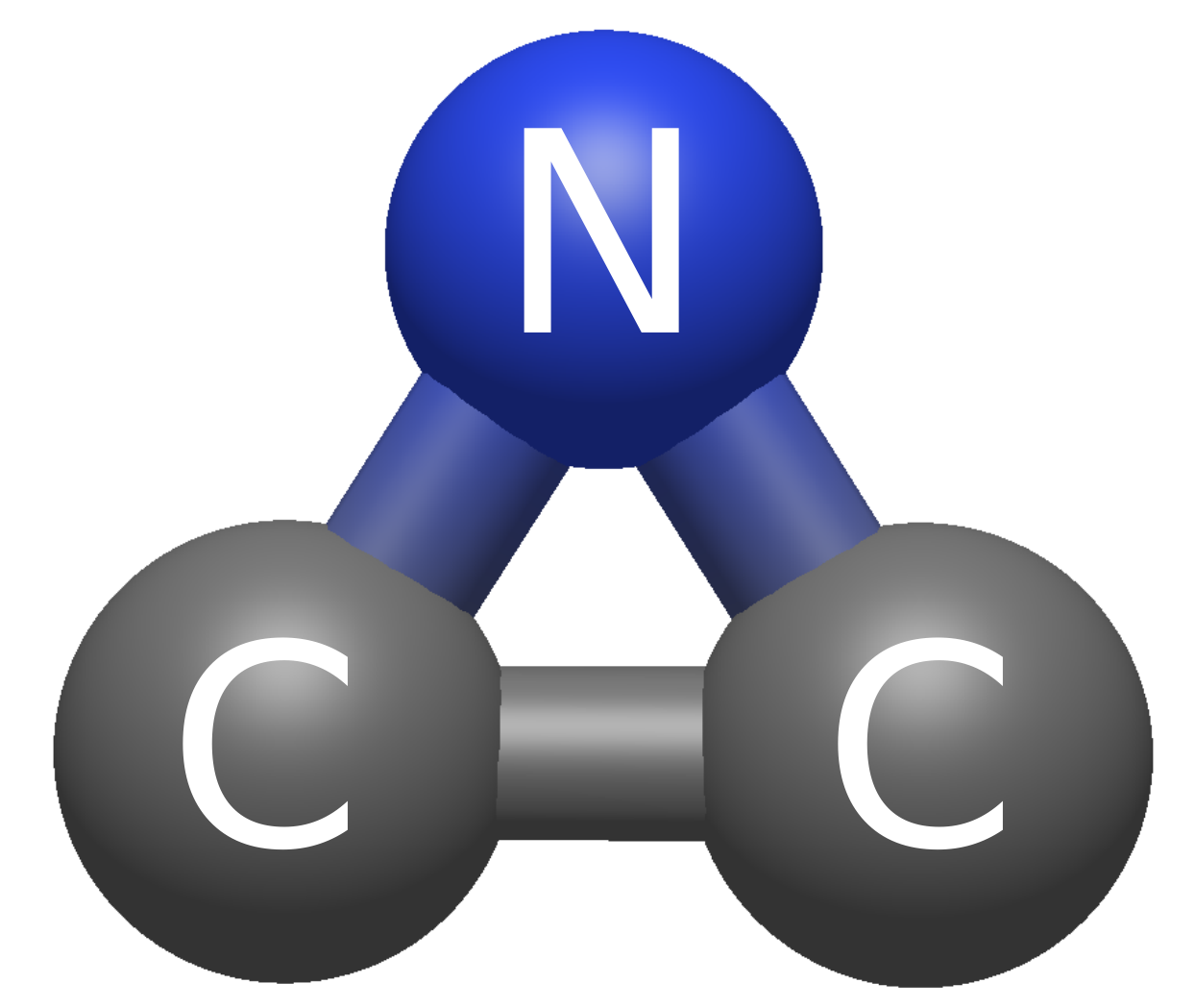}}} & \multicolumn{2}{c}{$c$-CNC$^{-}$} & & \multicolumn{2}{c}{$c$-$^{13}$CNC$^{-}$} & & \multicolumn{2}{c}{$c$-C$^{15}$NC$^{-}$} \\
\cline{2-3}\cline{5-6}\cline{8-9}\\[-2.25ex]
  & {QFF\tnote{a}} & {QFF\tnote{b}} & & {QFF\tnote{a}} & {QFF\tnote{b}} & & {QFF\tnote{a}} & {QFF\tnote{b}}\\[0.5ex] 
\hline \\[-2.25ex]
$r_{0}(\mathrm{N\!-\!C})$/\si{\pico\metre}        &   136.3000 &   136.2734 & &   136.2858 &   136.2593 & &   136.2903 &   136.2637 \\
                                                  &            &            & &   136.3007 &            & &            &            \\
$\angle_{0}(\mathrm{C\!-\!N\!-\!C})$/\si{\degree} &    64.800  &    64.812  & &    64.796  &    64.808  & &     64.804 &    64.816  \\
$A_{0}$                                           & 43384.08   & 43406.38   & & 42962.76   & 42983.82   & & 41567.27   & 41588.32   \\
$B_{0}$                                           & 39690.76   & 39692.55   & & 37968.13   & 37970.92   & & 39690.61   & 39692.89   \\
$C_{0}$                                           & 20644.40   & 20650.27   & & 20075.91   & 20081.47   & & 20222.86   & 20228.62   \\
$A_{1}$                                           & 43112.40   & 43135.49   & & 42683.37   & 42705.39   & & 41323.55   & 41345.21   \\
$B_{1}$                                           & 39544.95   & 39548.05   & & 37843.00   & 37846.70   & & 39532.99   & 39536.49   \\
$C_{1}$                                           & 20543.71   & 20550.14   & & 19980.79   & 19986.89   & & 20123.93   & 20130.19   \\
$A_{2}$                                           & 43682.21   & 43702.28   & & 43226.36   & 43245.79   & & 41841.29   & 41860.34   \\
$B_{2}$                                           & 39316.60   & 39322.04   & & 37639.10   & 37644.64   & & 39329.04   & 39334.66   \\
$C_{2}$                                           & 20526.23   & 20532.55   & & 19960.63   & 19966.63   & & 20111.50   & 20117.66   \\
$A_{3}$                                           & 43068.78   & 43092.86   & & 42682.84   & 42705.30   & & 41268.88   & 41291.64   \\
$B_{3}$                                           & 39775.85   & 39778.95   & & 38023.34   & 38027.50   & & 39774.62   & 39777.98   \\
$C_{3}$                                           & 20512.25   & 20519.05   & & 19949.32   & 19955.76   & & 20093.81   & 20100.50   \\
$10^{3}{\Delta}_{J}$                              &   117.187  &   117.065  & &   106.299  &   106.161  & &   116.772  &   116.650  \\
$10^{3}{\Delta}_{JK}$                             &   -80.725  &   -81.807  & &   -56.421  &   -57.192  & &   -90.260  &   -91.279  \\
$10^{3}{\Delta}_{K}$                              &   206.300  &   207.050  & &   186.291  &   186.687  & &   196.297  &   197.022  \\
$10^{3}{\delta}_{J}$                              &    49.439  &    49.381  & &    44.453  &    44.386  & &    49.647  &    49.589  \\
$10^{3}{\delta}_{K}$                              &    86.773  &    86.145  & &    90.374  &    89.849  & &    76.408  &    75.819  \\
$10^{6}{\varPhi}_{J}$                             &     0.091  &     0.101  & &     0.214  &     0.219  & &     0.091  &     0.101  \\
$10^{6}{\varPhi}_{JK}$                            &     5.884  &     5.802  & &     2.367  &     2.370  & &     5.475  &     5.395  \\
$10^{6}{\varPhi}_{KJ}$                            &   -23.461  &   -23.196  & &   -12.098  &   -12.081  & &   -21.689  &   -21.439  \\
$10^{6}{\varPhi}_{K}$                             &    19.839  &    19.635  & &    11.194  &    11.172  & &    18.191  &    18.002  \\
$10^{6}{\phi}_{J}$                                &     0.044  &     0.049  & &     0.106  &     0.108  & &     0.044  &     0.049  \\
$10^{6}{\phi}_{JK}$                               &     2.240  &     2.216  & &     0.780  &     0.790  & &     2.104  &     2.080  \\
$10^{6}{\phi}_{K}$                                &    -0.384  &    -0.341  & &     2.431  &     2.410  & &    -0.749  &    -0.706  \\
$\alpha^{A}_1$                                    &   271.7    &            & &   279.4    &            & &   243.7    &            \\
$\alpha^{A}_2$                                    &  -298.1    &            & &  -263.6    &            & &  -274.0    &            \\
$\alpha^{A}_3$                                    &   315.3    &            & &   279.9    &            & &   298.4    &            \\
$\alpha^{B}_1$                                    &   145.8    &            & &   125.1    &            & &   157.6    &            \\
$\alpha^{B}_2$                                    &   374.2    &            & &   329.0    &            & &   361.6    &            \\
$\alpha^{B}_3$                                    &   -85.1    &            & &   -55.2    &            & &   -84.0    &            \\
$\alpha^{C}_1$                                    &   100.7    &            & &    95.1    &            & &    98.9    &            \\
$\alpha^{C}_2$                                    &   118.2    &            & &   115.3    &            & &   111.4    &            \\
$\alpha^{C}_3$                                    &   132.2    &            & &   126.6    &            & &   129.0    &            \\
$r_{e}(\mathrm{N\!-\!C})$/\si{\pico\metre}        &   135.6664 &   135.6422 & &     {-}    &      {-}   & &      {-}   &     {-}    \\
$\angle_{e}(\mathrm{C\!-\!N\!-\!C})$/\si{\degree} &    64.746  &    64.759  & &     {-}    &      {-}   & &      {-}   &     {-}    \\
$A_{e}$                                           & 43528.50   & 43550.61   & & 43110.62   & 43131.31   & & 41701.32   & 41722.20   \\
$B_{e}$                                           & 39908.19   & 39907.41   & & 38167.60   & 38167.89   & & 39908.19   & 39907.66   \\
$C_{e}$                                           & 20819.90   & 20824.75   & & 20244.40   & 20249.04   & & 20392.53   & 20397.38   \\
$\mu_{e}$/$D$\tnote{c}                            &     1.1011 &     1.10   & &     1.0536 &            & &    1.1908  &            \\
\hline\\[-2.25ex]
$eQq(\mathrm{^{14}N})$\tnote{d}                   &     5.0986 &            & &    {-}     &            & &            &            \\
$\eta(\mathrm{^{14}N})$\tnote{d}                  &     0.0385 &            & &    {-}     &            & &            &            \\
\hline\hline
\end{tabular}
\begin{tablenotes}[flushleft]
  \item[a]{{\footnotesize This work. Data obtained using \texttt{SPECTRO}~\citep{SPECTRO}. Rotational constants ($A_{i}$, $B_{i}$, $C_{i}$), vibration-rotation interaction constants ($\alpha^{A}_{i}$, $\alpha^{B}_{i}$, $\alpha^{C}_{i}$), quartic (${\Delta}_{J}$, ${\Delta}_{JK}$, ${\Delta}_{K}$, ${\delta}_{J}$, ${\delta}_{K}$) and sextic (${\varPhi}_{J}$, ${\varPhi}_{JK}$, ${\varPhi}_{KJ}$, ${\varPhi}_{K}$, ${\phi}_{J}$, ${\phi}_{JK}$, ${\phi}_{K}$) centrifugal distortion parameters are all in MHz. {$A_{i}$, $B_{i}$, $C_{i}$ with $i\!=\!1$-$3$ are effective rotational constants calculated for the three vibrational fundamentals; see Figure~\ref{fig:convergence} for mode descriptions. The corresponding zero-point level constants are $A_{0}$, $B_{0}$, $C_{0}$}}.}
  \item[b]{{\footnotesize Ref.~\citenum{FOR017:22860}}.}
  \item[c]{{\footnotesize CBS-extrapolated 
  dipole moments (in D) at the QFF equilibrium geometry; see Ref.~\citenum{CON020:024105}}.}
  \item[d]{{\footnotesize CBS-extrapolated 
  $\mathrm{^{14}N}$ nuclear quadrupole (hyperfine) coupling constant ($eQq$ in MHz) and asymmetry parameter ($\eta$  unitless) at the QFF equilibrium geometry; see text}.}  
\end{tablenotes}
\end{threeparttable}
\end{table*}

A close inspection of Figure~\ref{fig:convergence} unravels the slow convergence rates of the raw CC/ACV$X$Z ($X\!=\!T,Q,5$) harmonic frequencies towards the predicted CBS values, 
a feature clearly expected from standard all electron (ae) CC methods~\cite{TAJ004:11599}; see shaded gray areas and the black dashed lines outlined therein. A similar convergence behavior is also found for the $R_{i}^{e}$'s (Figure~S1).   
Indeed, on going from the smaller ACV$T$Z to the ACV$5$Z basis set, the predicted CCSD(T) 
equilibrium bond distances decrease by $\approx\!0.01\,\si{\angstrom}$, followed by increments of $\approx\!10\,\mathrm{cm^{-1}}$ in the corresponding harmonic frequencies. In turn, extrapolations to the CBS limit produce only minor changes in the 
CC/ACV$5$Z attributes, as expected: the $\omega_{i}$'s 
and $R_{i}^{e}$'s vary by less than $+2\,\mathrm{cm^{-1}}$ and $-0.002\,\si{\angstrom}$/$-0.005\,\mathrm{deg}$, respectively. Moreover, the inclusion of scalar relativity [$\Delta_{\mathrm{DKH}}$ in Eq.~(\ref{eq:rel})] into $E^{\mathrm{CC}}_{\infty}$ [Eqs.~(\ref{eq:cbsenergy})-(\ref{eq:cbscor})] leads to only slight, but still significant reductions in both $\omega_{i}$ and $R_{i}^{e}$. As expected for such light molecules, these corrections are small and amount to $\approx\!-62\,\mathrm{mE_{h}}$ around equilibrium. In addition to the one-particle basis set correlation recovery  
at CC, Figures~\ref{fig:convergence} and~S1 also permit an approximate assessment of the $\omega_{i}$ and $R_{i}^{e}$ convergence rates upon increasing coupled cluster $\mathcal{N}$-particle expansions~[CCSD(T)$\rightarrow$CCSDT$\rightarrow$\!CCSDTQ$\longrightarrow$$\approx$FCI]; see white regions and orange dashed lines. When added to the $E^{\mathrm{CC}}_{\infty}\!+\!\Delta_{\mathrm{DKH}}$ components, the CCSDT corrections to CCSD(T), $\Delta\mathrm{T}$ in Eq.~(\ref{eq:ho}), 
are shown to slightly 
overestimate (underestimate) the bond angle (bond distances) of 
$c$-CNC$^{-}$($^{1}A_{1}$), thus leading to a net increase in the $\omega_{i}$'s; the contrary is the case for the linear species. Indeed, the $\Delta\mathrm{T}$ correlation contributions to the total energies 
assume opposing signs for $c$-CNC$^{-}$($^{1}A_{1}$) and $\ell$-CCN$^{-}$($^{3}\Sigma^{-}$), being of the order of $+0.6$ and $-1\,\mathrm{mE_{h}}$, respectively. 
Yet, faster convergence rates of these equilibrium properties 
towards their (approximate) 
FCI limit values are 
clearly perceived after the addition of the correlation component due to iterative quadruples, 
$\Delta\mathrm{Q}$ [Eq.~(\ref{eq:ho})]. As noted elsewhere~\citep{FEL008:204105,KAR010:144102,KAR020:024102}, the weaker dependence on the basis set size and the increasingly 
faster convergence rates of the HO terms (notably, $\Delta\mathrm{Q}$ and corrections beyond it) 
are demonstrably attributed 
to their intimate relation to nondynamical 
rather than dynamical correlation. In fact, at the CCSDTQ level, the predicted equilibrium geometries and harmonic frequencies appear to be well converged for both species, with the inclusion 
of the $\Delta\mathrm{FCI}$ [Eqs.~(\ref{eq:ho})~and~(\ref{eq:cf})] terms being 
responsible for less than $0.0001\,\si{\angstrom}$/$-0.005\,\mathrm{deg}$ and $-1\,\mathrm{cm^{-1}}$ adjustments, respectively; see Figures~\ref{fig:convergence} and~S1. As expected, $\Delta\mathrm{Q}$ is 
the largest among all HO corrections, amounting to $\approx\!-1.5\,\mathrm{mE_{h}}$; $\Delta\mathrm{FCI}$ is of the order of $-0.1\,\mathrm{mE_{h}}$. 
To gauge the reliability of our predicted $\Delta\mathrm{FCI}$ corrections to CCSDTQ, we have followed Ref.~\citenum{FEL008:204105} and for comparison estimated this limit at the QFF equilibrium geometries with the cf approximant [Eq.~(\ref{eq:cf})] but using instead the CCSDT/V$D$Z, CCSDTQ/V$D$Z, and
CCSDTQP/V$D$Z higher hierarchical sequence; the final FCI corrections obtained in this way include additional $\Delta P$ terms and are thereafter denoted as $\Delta\mathrm{FCI}_{\mathrm{true}}$. 
The results have shown {\color{black}that our} ``cost-effective'' cf protocol based on the CCSD$\rightarrow$CCSDT$\rightarrow$\!CCSDTQ sequence series recovers nearly $80\%$ of the ``true`` residual FCI correlation energies [\emph{i.e.}, $\Delta\mathrm{FCI}$/$\Delta\mathrm{FCI}_{\mathrm{true}}\!\approx\!0.75$ and $0.8$ for $\ell$-CCN$^{-}$($^{3}\Sigma^{-}$) and $c$-CNC$^{-}$($^{1}A_{1}$), respectively], {hence} confirming its feasibility and accuracy. Similar conclusions were drawn in previous studies~\citep{FEL008:204105,FEL006:054107}. 
It should be emphasized that, although specialized extrapolation formulas have recently been developed for HO terms~\citep{KAR020:024102}, no attempts have here been made 
to estimate these corrections at the CBS limit as this would imply a formidable 
computational effort. Thus, small additional (residual) one-particle basis set 
truncation errors may still be foreseen in our protocol, albeit with conceivably little impact on the final results. Of course, the
magnitude of these uncertainties {\color{black}is} expected to 
exceed those effects associated with diagonal Born-Oppenheimer (DBOC)  
and non-adiabatic corrections~\citep{REI015:24641}, and we therefore opted not to include them either in our
approach. Note that a high-level QFF already exists in the literature for $c$-CNC$^{-}$($^{1}A_{1}$)~\citep{FOR017:22860}. It is based 
on the well-established CcCR protocol which includes, in addition to CBS-extrapolated CC/AV$X$Z ($X\!=\!T,Q,5$) energies, 
core-valence and relativistic effects~\citep{XIN009:104301,HUA008:044312,FOR015:9941,GAR021:119184}. For comparison, we also plot in Figures~\ref{fig:convergence}~(b)~and~S1~(b) the predicted CcCR equilibrium {\color{black}properties taken from Ref.~\citenum{FOR017:22860}}. Accordingly, the CcCR results agree quite well with our predicted attributes, notably when compared with the $E^{\mathrm{CC}}_{\infty}\!+\!\Delta_{\mathrm{DKH}}$ PES (as expected). This provides further evidence on the reliability of our CBS extrapolation protocol [Eqs.~(\ref{eq:cbsenergy})-(\ref{eq:cbscor})]. Yet, the small discrepancies 
found for $R_{i}^{e}$ and $\omega_{i}$ ($\approx\!0.0002\,\si{\angstrom}$/$0.01\,\mathrm{deg}$ and $\lesssim\!4\,\mathrm{cm^{-1}}$) are clearly attributed to the effects of HO correlations; see 
Tables~\ref{tab:force},~\ref{tab:rotcyc}~,~\ref{tab:vibcyc} and later discussions for further comparisons. {\color{black}Unfortunately, no experimental data is yet available for 
this species. This is due to the combined fact that $\mathrm{C_{2}N^{-}}$ molecules are not easy to produce in larger amounts and may be spectroscopically hard to identify without guiding theoretical predictions.} 

{\color{black}\subsection{Molecular structures and rotational constants}\label{equirot}}
As Table~\ref{tab:rotlin} shows, the computed $\mathrm{C\!-\!C}$ and $\mathrm{C\!-\!N}$ equilibrium bond distances for $\ell$-CCN$^{-}$($^{3}\Sigma^{-}$) 
using our final composite PES 
are $1.359387$~and~$1.203706\,\si{\angstrom}$, respectively. This structure is fairly close to the best-guess 
initial geometry ($1.359176$ and $1.201802\,\si{\angstrom}$) utilized in the generation of the QFF \emph{ab initio} grid points (section~\ref{sec:abinitio}). 
To the best of our knowledge, available literature data on the $\ell$-CCN$^{-}$($^{3}\Sigma^{-}$) equilibrium attributes is somewhat limited to the \emph{ab initio} B3LYP studies by Pascoli~\citep{PAS99:333} 
and Garand~\emph{et al.}~\citep{GAR009:064304} who reported 
$r_{e}(\mathrm{C\!-\!C})/r_{e}(\mathrm{C\!-\!N})\!=\!1.344\,\si{\angstrom}/1.207\,\si{\angstrom}$. Recent MRCI(Q)/AV$Q$Z calculations by Franz~\emph{et al.}~\citep{FRA020:234303} put better constraints on these values ($1.360$~and~$1.212\,\si{\angstrom}$). 
Indeed, our predicted equilibrium rotational constant, $B_{e}$, of $11903.44\,\mathrm{MHz}$ for the main isotopologue 
is $\approx\!1\%$ lower and greater than the corresponding DFT~\citep{PAS99:333,GAR009:064304} ($12015.49\,\mathrm{MHz}$) and MRCI~\citep{FRA020:234303} ($11820.70\,\mathrm{MHz}$) values, respectively, and is expected to be the most reliable theoretical estimate 
currently available. With inclusion 
of vibrational (zero-point level) corrections 
via the VPT2-based $\alpha^{B}_{i}$'s, $B_{e}$ is found to decrease by $41.52\,\mathrm{MHz}$ 
($B_{0}\!=\!11861.91\,\mathrm{MHz}$), consistent with 
an increase in the vibrationally-averaged $r_{0}$ bond distances; similar trends follow for the linear rare isotopologues (Table~\ref{tab:rotlin}).  
{\color{black}Note that, for $\ell$-$^{13}$CCN$^{-}$ and $\ell$-CC$^{15}$N$^{-}$, 
the calculated $B_{0}$ constants, $11369.30$ and $11490.06\,\mathrm{MHz}$, show large isotopic shifts (as expected), being $\approx\!430\,\mathrm{MHz}$ smaller on average than the predicted $B_{0}$ for $\ell$-CCN$^{-}$. 
Thus, differently from $\ell$-C$^{13}$CN$^{-}$ (see Table~\ref{tab:rotlin}), their pure rotational spectra should 
be clearly distinguished from that of the main isotopologue. {Also quoted in Table~\ref{tab:rotlin} are the associated effective rotational constants for the three vibrational fundamentals, $B_{i}$\,($i\!=\!1$-$3$); see also Figure~\ref{fig:convergence} for mode descriptions.} 

The corresponding spectroscopic attributes obtained for $c$-CNC$^{-}$($^{1}A_{1}$) using our best composite QFF and VPT2 
are presented in Table~\ref{tab:rotcyc}, wherein the most accurate results from the literature~\citep{FOR017:22860} are also listed for comparison. Accordingly, the predicted equilibrium $\mathrm{N\!-\!C}$ bond distance and $\mathrm{C\!-\!N\!-\!C}$ angle are $1.356664\,\si{\angstrom}$ and $64.746\,\si{\degree}$, respectively. Again, these values are quite close (as expected) to the ones employed as 
starting reference geometry ($1.355716\,\si{\angstrom}$ and $64.789\,\si{\degree}$) and to those reported 
from the CcCR QFF ($1.356422\,\si{\angstrom}$ and $64.759\,\si{\degree}$)~\citep{FOR017:22860}. As clearly perceived from Table~\ref{tab:rotcyc}, $c$-CNC$^{-}$($^{1}A_{1}$) is representative of a (near-oblate) asymmetric top 
with $A_{e}\!=\!43528.50$, $B_{e}\!=\!39908.19$ and $C_{e}\!=\!20819.90\,\mathrm{MHz}$, values that differ by less than $0.05\%$ from 
those predicted by the CcCR protocol~\citep{FOR017:22860}; the largest discrepancy (of up to $\sim\!20\,\mathrm{MHz}$) is found for $A_{e}$ as this appears to be the most sensitive to electron correlation~\citep{GAR021:119184}. 
Using the effective rotational constants for the zero-point level ($A_{0}$, $B_{0}$, and $C_{0}$), 
the calculated Ray's asymmetry parameters~\citep{GOR84:MMS,COO012:698392}, $\kappa$, for the main isotopologue are thus 
$0.675$ (this work) and $0.674$ (CcCR), {hence} further suggesting the nearly statistical equivalence of these 
two theoretical 
data sets. 
{\color{black} The corresponding $\kappa$ values calculated here   
for $c$-$^{13}$CNC$^{-}$ 
and $c$-C$^{15}$NC$^{-}$ 
are $0.563$ and $0.824$, respectively. So, as expected, substitution by $^{13}$C or $^{15}$N makes the 
corresponding 
ground-state vibrationally-averaged structures 
deviate further from or even closer to the oblate symmetric top limit ($\kappa\!=\!+1$~\citep{GOR84:MMS,COO012:698392}), respectively. As in the case of the linear form, 
clear differences should then be apparent in the pure rotational spectra of $c$-CNC$^{-}$ 
and its rare isotopologues. {For all these species and for future reference, we also collect in Table~\ref{tab:rotcyc} the theoretically-predicted effective rotational constants for the three vibrational fundamentals, $A_{i}$, $B_{i}$, and $C_{i}$ with $i\!=\!1$-$3$.} 
}
\begin{table}
\begin{small}
\centering
\caption{\footnotesize Calculated harmonic and anharmonic vibrational frequencies (in $\mathrm{cm^{-1}}$) for $\ell$-CCN$^{-}$($^{3}\Sigma^{-}$) isotopologues using our final composite QFF (Table~\ref{tab:force}) and \texttt{SPECTRO}/\texttt{DVR3D}.}
\label{tab:viblin}
\begin{threeparttable}
\begin{tabular}{
l
l
c
S[table-align-text-post=false,table-format=4.1]@{}
S[table-align-text-post=false,table-format=4.1]
S[table-align-text-post=false,table-format=4.1]
S[table-align-text-post=false,table-format=4.1]}
\hline\hline \\[-1.8ex]
 {Molecule} & {Description} & {Mode} & {Harmonic} & {VPT2\tnote{a}} & {VAR\tnote{b}} \\[0.5ex]
\hline \\[-2.25ex]
$\ell$-CCN$^{-}$        &  $\sigma\,\mathrm{C\!-\!N}$ stretch          & $\nu_{1}$                    & 1759.9 & 1695.6 & 1696.5       \\
                        &  $\pi\,\mathrm{C\!-\!C\!-\!N}$ bend          & $\nu_{2}$                    &  458.3 &  451.9 &  452.1       \\
                        &                                              &                              &        & \multicolumn{2}{c}{452.9$\pm$2.9\tnote{c}}  \\
                        &  $\sigma\,\mathrm{C\!-\!C}$ stretch          & $\nu_{3}$                    & 1055.5 & 1046.9 & 1045.8       \\
                        &  zero-point energy                           & {ZPE}                    & 1866.0 & 1851.0 & 1850.3       \\
\\ 
$\ell$-$^{13}$CCN$^{-}$ &                                              & $\nu_{1}$                    & 1759.4 & 1694.6 & 1695.6       \\
                        &                                              & $\nu_{2}$                    &  455.6 &  449.4 &  449.6       \\
                        &                                              & $\nu_{3}$                    & 1027.5 & 1021.7 & 1020.6       \\
                        &                                              & {ZPE}                    & 1849.1 & 1834.2 & 1833.6       \\
\\                                                                                                         
$\ell$-C$^{13}$CN$^{-}$ &                                              & $\nu_{1}$                    & 1719.4 & 1659.1 & 1659.9       \\
                        &                                              & $\nu_{2}$                    &  446.1 &  440.0 &  440.2       \\
                        &                                              & $\nu_{3}$                    & 1051.4 & 1040.1 & 1039.2       \\
                        &                                              & {ZPE}                    & 1831.5 & 1817.1 & 1816.4       \\
\\
$\ell$-CC$^{15}$N$^{-}$ &                                             & $\nu_{1}$                    & 1736.5 & 1673.2 & 1674.2       \\
                        &                                             & $\nu_{2}$                    &  455.8 &  449.5 &  449.7       \\
                        &                                             & $\nu_{3}$                    & 1047.0 & 1039.5 & 1038.4       \\
                        &                                             & {ZPE}                    & 1847.6 & 1832.8 & 1832.2       \\
\hline\hline
\end{tabular}
\begin{tablenotes}[flushleft]
  \item[a]{{\footnotesize This work. Data obtained using \texttt{SPECTRO}~\citep{SPECTRO} and 
  the internal-coordinate force field.}}
  \item[b]{{\footnotesize This work. Data obtained using the variational (VAR), exact kinetic energy nuclear motion code \texttt{DVR3D}~\citep{TEN004:85} and the QFF transformed into a Morse-sine coordinate system. 
  }}
  \item[c]{{\footnotesize Experimental value derived from Refs.~\citenum{GAR009:064304}~and~\citenum{MUZ015:105}.}}
\end{tablenotes}
\end{threeparttable}
\end{small}
\end{table}

Also listed in Tables~\ref{tab:rotlin}~and~\ref{tab:rotcyc} are the calculated 
dipole moments, $\mu_{e}$, at the QFF equilibrium geometries for the various isotopologues; these were obtained using CC/ACV$X$Z ($X\!=\!Q,5$) energies and the CBS extrapolation protocol of Ref.~\citenum{CON020:024105}. 
Note that, {with the exception of $c$-$^{13}$CNC$^{-}$}, all dipoles are oriented along the negative z-axis, with the negative charge located on the N atom; the corresponding origins lie at the isotopologues' center-of-mass. Indeed, the large $\mu_{e}$ values so found, particularly for the linear forms ($\approx\!2.0\,\mathrm{D}$), indicate that these anions might be fairly 
bright for GBT and ALMA (see later section~\ref{sec:astro}), provided the abundance of a particular isotopologue 
is large enough to be detectable. 
\begin{table*}[htb!]
\centering
\caption{\footnotesize Calculated harmonic and anharmonic vibrational frequencies (in $\mathrm{cm^{-1}}$) for $c$-CNC$^{-}$($^{1}A_{1}$) isotopologues using our final composite QFF (Table~\ref{tab:force}) and \texttt{SPECTRO}/\texttt{DVR3D}.}
\label{tab:vibcyc}
\begin{threeparttable}
\begin{tabular}{
l
l
c
S[table-align-text-post=false,table-format=4.1]
S[table-align-text-post=false,table-format=4.1]
S[table-align-text-post=false,table-format=4.1]
S[table-align-text-post=false,table-format=4.1]}
\hline\hline \\[-1.8ex]
 {Molecule} & {Description} & {Mode} & {Harmonic} & {VPT2\tnote{a}} & {VPT2\tnote{b}} & {VAR\tnote{c}} \\[0.5ex]
\hline \\[-2.25ex]                                                                                                                     
$c$-CNC$^{-}$    &  $a_{1}\,\mathrm{C\!-\!N}$ symm. stretch     & $\nu_{1}$                    & 1493.7 & 1461.5 & 1462.9 &      1460.3 \\
                 &  $a_{1}\,\mathrm{C\!-\!N\!-\!C}$ bend        & $\nu_{2}$                    & 1011.9 &  990.2 &  991.7 &       993.6 \\
                 &  $b_{2}\,\mathrm{C\!-\!N}$ antisymm. stretch & $\nu_{3}$                    & 1068.8 & 1045.5 & 1049.4 &      1046.4 \\
                 &  zero-point energy                           & {ZPE}                    & 1787.2 & 1777.5 & 1780.6 &      1778.1 \\
\\                                                                                                                                          
$c$-$^{13}$CNC$^{-}$ &                                          & $\nu_{1}$                    & 1476.1 & 1444.7 & 1446.2 &      1443.5 \\
                 &                                              & $\nu_{2}$                    &  997.1 &  976.0 &  977.5 &       979.3 \\
                 &                                              & $\nu_{3}$                    & 1054.3 & 1031.6 & 1035.5 &      1032.5 \\
                 &                                              & {ZPE}                    & 1763.8 & 1754.3 & 1757.4 &      1754.9 \\
\\                                                                                                                                          
$c$-C$^{15}$NC$^{-}$ &                                          & $\nu_{1}$                    & 1475.1 & 1443.6 & 1445.0 &      1442.6 \\
                 &                                              & $\nu_{2}$                    & 1002.9 &  981.6 &  983.1 &       984.9 \\
                 &                                              & $\nu_{3}$                    & 1057.0 & 1034.3 & 1038.1 &      1035.2 \\
                 &                                              & {ZPE}                    & 1767.5 & 1758.0 & 1761.1 &      1758.6 \\
\hline\hline
\end{tabular}
\begin{tablenotes}[flushleft]
  \item[a]{{\footnotesize This work. Data obtained using \texttt{SPECTRO}~\citep{SPECTRO} 
  and the internal-coordinate force field.}}
  \item[b]{{\footnotesize Ref.~\citenum{FOR017:22860}}.}
  \item[c]{{\footnotesize This work. Data obtained using the variational (VAR), exact kinetic energy nuclear motion code \texttt{DVR3D}~\citep{TEN004:85} and the QFF transformed into a Morse-cosine coordinate system.}}
\end{tablenotes}
\end{threeparttable}
\end{table*}
\begin{table*}[htb!]
\centering
\caption{\footnotesize Anharmonic constants (in $\mathrm{cm^{-1}}$) of $\ell$-CCN$^{-}$($^{3}\Sigma^{-}$) and $c$-CNC$^{-}$($^{1}A_{1}$) isotopologues.~$^{\rm a}$}
\label{tab:anharm}
\begin{threeparttable}
\begin{tabular}{
l
S[table-align-text-post=false,table-format=4.3]
S[table-align-text-post=false,table-format=4.3]
S[table-align-text-post=false,table-format=4.3]
S[table-align-text-post=false,table-format=4.3]
S[table-align-text-post=false,table-format=4.3]
S[table-align-text-post=false,table-format=4.3]
S[table-align-text-post=false,table-format=4.3]}
\hline\hline \\[-1.8ex]
  & {$\ell$-CCN$^{-}$} & {$\ell$-$^{13}$CCN$^{-}$} & {$\ell$-C$^{13}$CN$^{-}$} & {$\ell$-CC$^{15}$N$^{-}$} & {$c$-CNC$^{-}$} & {$c$-$^{13}$CNC$^{-}$} & {$c$-C$^{15}$NC$^{-}$} \\
\hline \\[-2.25ex]
$x_{11}$ & -27.000 & -27.016 & -25.866 & -26.135 &  -5.225 &  -5.200 &  -5.015 \\
$x_{12}$ & -10.483 & -10.371 &  -9.675 & -10.585 & -16.993 & -16.125 & -17.421 \\
$x_{13}$ &   0.277 &  -0.783 &   2.233 &  -0.850 & -26.525 & -25.918 & -25.574 \\
$x_{22}$ &   0.996 &   0.954 &   0.910 &   1.042 &  -4.757 &  -4.752 &  -4.443 \\
$x_{23}$ &  -7.004 &  -6.735 &  -6.832 &  -6.912 &  -7.369 &  -6.984 &  -7.473 \\
$x_{33}$ &  -8.489 &  -7.724 &  -8.986 &  -8.068 &  -3.193 &  -3.134 &  -3.126 \\
$g_{22}$ &  -0.623 &  -0.582 &  -0.557 &  -0.674 &         &         &         \\
\hline\hline
\end{tabular}
\begin{tablenotes}[flushleft]
  \item[a]{{\footnotesize This work. Data determined from 
  our final composite internal-coordinate QFFs (Table~\ref{tab:force}) 
  using second-order perturbation theory~\citep{NIE51:90,MIL72:115,ALI85:1} 
  as implemented in \texttt{SPECTRO}~\citep{SPECTRO}}.}
\end{tablenotes}
\end{threeparttable}
\end{table*}

{\color{black}\subsection{(Hyper)fine splittings}\label{hyperfine}}
To aid in future high-resolution laboratory investigations on C$_{2}$N$^{-}$, we also provide in Tables~\ref{tab:rotlin}~and~\ref{tab:rotcyc} reliable estimates of fine and 
hyperfine 
coupling constants~\citep{PUZ010:273}; see also Table~S5. 
For $\ell$-CCN$^{-}$($^{3}\Sigma^{-}$), the calculation of the relevant spin-spin coupling at the QFF equilibrium geometry, $\lambda_{e}\!=\!\lambda^{e}_{\mathrm{SO}}\!+\!\lambda^{e}_{\mathrm{SS}}$, followed 
the formalism of Vahtras~\emph{et al.}~\citep{VAH002:133} where $\lambda^{e}_{\mathrm{SO}}$ is 
the contribution due to second-order spin-orbit (SO) effects, while $\lambda^{e}_{\mathrm{SS}}$ describes the    
magnetic dipole-dipole (electron) spin-spin (SS) interactions~\citep{VAH002:133}. 
In turn, the estimation of the (electron) spin-rotation coupling constant ($\gamma_{e}$) relied solely~\citep{TAR010:9246} on 
its approximate relation to the electronic $g$-tensor as derived by Curl~\citep{CUR65:585}, $\gamma_{e}\!=\!-2B_{e}\Delta g_{\perp}$, where $\Delta g_{\perp}$ is the transversal component of the calculated $g$-shift~\citep{ENG98:149} and $B_{e}$ the rotational constant; all such fine structure attributes were herein obtained at the full-valence CASSCF/AV$Q$Z level of theory using \texttt{DALTON} software suite~\citep{DALTON}. To assess the reliability of such an approach, we have applied it to the isoelectronic $\ell$-$\mathrm{CCO}(^{3}\Sigma^{-})$ species for which accurate experimental $\lambda_{0}$ and $\gamma_{0}$ values are available~\citep{ABU006:206}. 
The calculated constants, $\lambda_{e}\!=\!11594.05\,\mathrm{MHz}$ and $\gamma_{e}\!=\!-15.04\,\mathrm{MHz}$, are in excellent agreement with the observed values~\citep{ABU006:206}, $11600$ and $-17.82\,\mathrm{MHz}$, and this is the accuracy one might expect for $\ell$-CCN$^{-}$($^{3}\Sigma^{-}$); {\color{black} note here the expectedly small dependence of $\lambda$ and $\gamma$ on zero-point vibrational 
corrections~\citep{ABU006:206,CAZ016:A126}.} 
As noted previously for $\mathrm{O_{2}}(^{3}\Sigma^{-}_{g})$~\citep{VAH002:133}, 
$\ell$-$\mathrm{CCO}(^{3}\Sigma^{-})$~\citep{ABU006:206} and 
actually the case here, 
$\lambda_{e/0}$ is of the order of $B_{e/0}$; see Table~\ref{tab:rotlin}. 

The (hyperfine) interaction between the $\mathrm{^{14}N}$ electric quadrupole moment [$Q(\mathrm{^{14}N})\!=\!0.02044\,\mathrm{barn}$~\citep{PYY018:1328}] and the molecular electric field gradient at $\mathrm{^{14}N}$ 
(EFG with principal-axis components $|\mathcal{V}_{zz}(\mathrm{^{14}N})|\!>\!|\mathcal{V}_{yy}(\mathrm{^{14}N})|\!>\!|\mathcal{V}_{xx}(\mathrm{^{14}N})|$) 
is herein 
defined by two additional parameters: the nuclear quadrupole coupling constant, $eQq(\mathrm{^{14}N})\!\propto\!
Q(\mathrm{^{14}N})\mathcal{V}_{zz}(\mathrm{^{14}N})$, and the asymmetry parameter, $\eta(\mathrm{^{14}N})\!=\![\mathcal{V}_{xx}(\mathrm{^{14}N})\!-\!\mathcal{V}_{yy}(\mathrm{^{14}N})]/\mathcal{V}_{zz}(\mathrm{^{14}N})$~\citep{LEH73:79,ANT019:224302,LIN98:188}; see Tables~\ref{tab:rotlin}~and~\ref{tab:rotcyc}. 
Note that, similarly to $\mu_{e}$, the relevant EFG tensors   
have been computed via finite field calculations in \texttt{MOLPRO}, with the corresponding CC/ACV$X$Z ($X\!=\!Q,5$) raw energies being likewise extrapolated to the CBS limit prior to the energy derivative evaluations~\citep{CON020:024105}. Again, for benchmark purposes, $\mathrm{^{14}N}$ hyperfine parameters were derived for the parent $\mathrm{CN}^{-}(^{1}\Sigma)$ 
and $\ell$-$\mathrm{C_{3}N}^{-}(^{1}\Sigma)$ species using the above protocol  
and compared with available experimental data~\citep{GOT007:191101,THA008:1132}. The calculated (observed) $eQq(\mathrm{^{14}N})$'s are $-4.265\,\mathrm{MHz}$ ($-4.238\pm0.032\,\mathrm{MHz}$~\citep{GOT007:191101}) and $-3.253\,\mathrm{MHz}$ ($-3.248\pm0.005\,\mathrm{MHz}$~\citep{THA008:1132}), respectively; our theoretical {\color{black}values}   
for $\ell$-CCN$^{-}$($^{3}\Sigma^{-}$) and $c$-CNC$^{-}$($^{1}A_{1}$) 
are $-2.909$ and $5.099\,\mathrm{MHz}$. 
Note that, for $\ell$-CCN$^{-}$($^{3}\Sigma^{-}$), 
an additional effect arises due to the intrinsic magnetic (hyperfine) interactions between nuclei with nonzero spin 
(\emph{e.g.}, $\mathrm{^{14}N}$, $\mathrm{^{15}N}$, $\mathrm{^{13}C}$) 
and the spin of the unpaired electrons. Such an electron spin-nuclear spin coupling 
has been shown to dominate the observed hyperfine structure of small open-shell $\Sigma$ species~\citep{FIT005:084312}. The corresponding magnetic hyperfine parameters, \emph{i.e.}, the isotropic Fermi contact ($b_{F}$) and anisotropic spin dipolar ($c$) interaction constants~\citep{FIT005:084312}, at specific nucleus are listed in Table~\ref{tab:rotlin}; 
these were computed at the full-valence CASSCF/ACV$5$Z level in \texttt{DALTON}~\citep{DALTON}.    
Suffice it to add that, for both $\ell$-CCN$^{-}$($^{3}\Sigma^{-}$) and $c$-CNC$^{-}$($^{1}A_{1}$), the contributions of other magnetic coupling tensors, \emph{e.g.}, nuclear spin-rotation, amount to only a few $\mathrm{kHz}$; for completeness, they are gathered in Table~S5. {\color{black}As emphasized later in section~\ref{sec:astro}, 
the consideration of all such (hyper)fine structure provides an additional 
spectroscopic identity to C$_{2}$N$^{-}$ that might be fairly handy for its unambiguous detection.}
\begin{figure*}
\captionsetup[subfigure]{labelformat=empty}
\centering
\subfloat{{\includegraphics[width=0.5\linewidth]{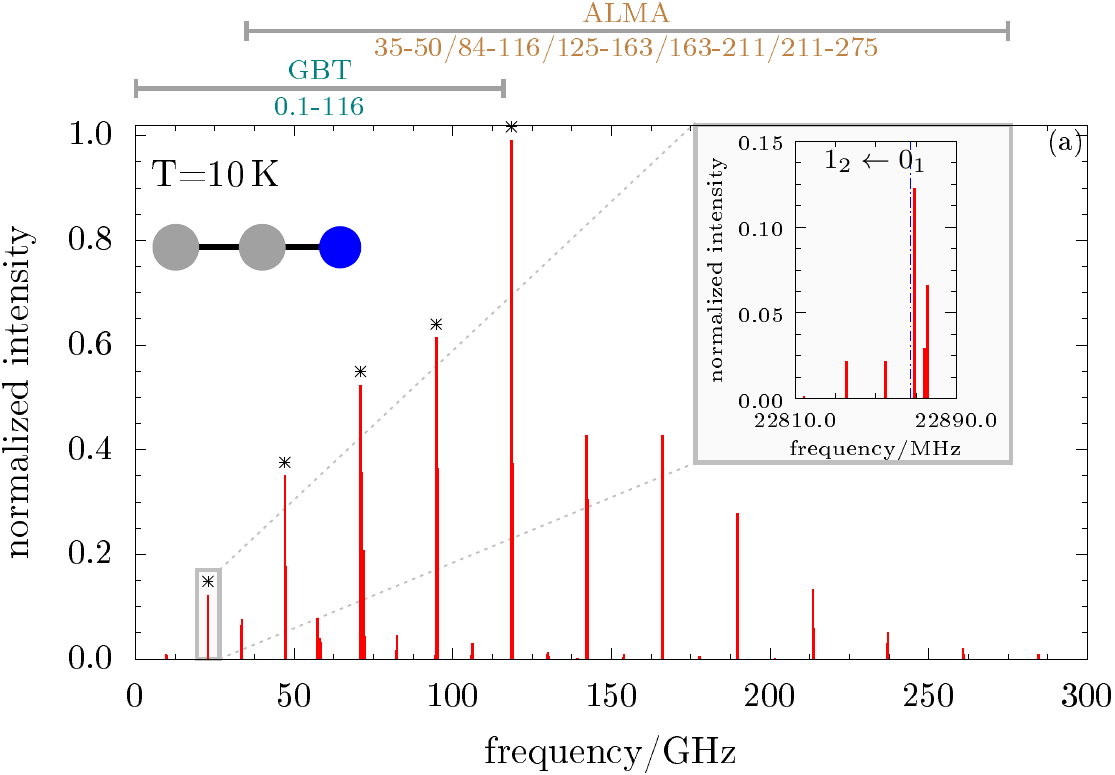}}}
\hfill
\subfloat{{\includegraphics[width=0.5\linewidth]{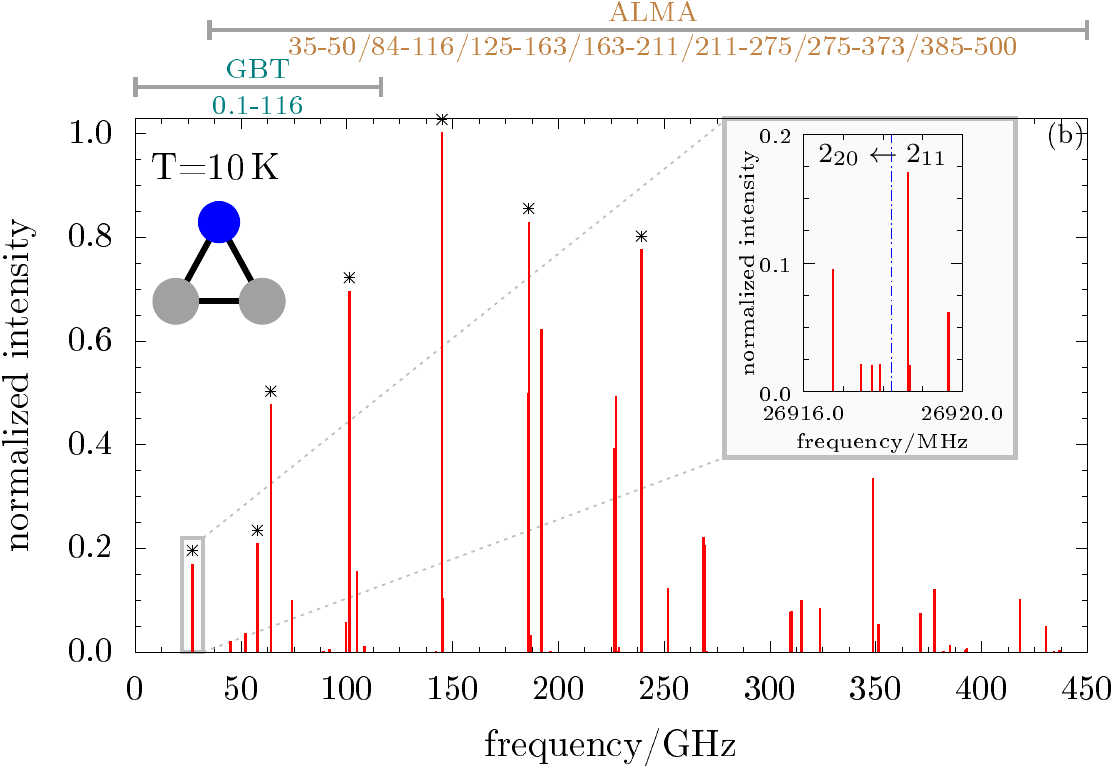}}}
\caption{\footnotesize Simulated rotational spectra at $10\,\mathrm{K}$ of~(a).~$\ell$-CCN$^{-}$($^{3}\Sigma^{-}$)~and~(b).~$c$-CNC$^{-}$($^{1}A_{1}$) in their ground vibrational states using \texttt{PGOPHER}~\citep{WES017:221} and the spectroscopic parameters presented in Tables~\ref{tab:rotlin},~\ref{tab:rotcyc},~and~S5: {\color{black}these are $A_{0}$, $B_{0}$ and $C_{0}$ for the rotational constants and equilibrium values for the centrifugal distortion, fine and hyperfine parameters (see text).} Working frequency ranges (in GHz) of the GBT and ALMA {[from left to right: band 1 (both panels), 
and bands 3-6 (panel~(a)) and bands 3-8 (panel~(b))]} are also displayed. The insets show the expected hyperfine splitting patterns for selected low-frequency transitions; blue dash-dotted lines highlight their positions  without consideration of hyperfine couplings. Lines marked with '*` are listed in Table~\ref{tab:lines} in increasing order of their frequencies.} 
\label{fig:spec}
\end{figure*}

{\color{black}\subsection{Fundamental vibrational frequencies}\label{fundfreq}}
Tables~\ref{tab:viblin}~and~\ref{tab:vibcyc} report the fundamental vibrational frequencies of $\ell$-CCN$^{-}$($^{3}\Sigma^{-}$) and $c$-CNC$^{-}$($^{1}A_{1}$) computed with VPT2/VAR and the composite force fields. Also shown for comparison are the corresponding harmonic frequencies and available results from the literature~\citep{FOR017:22860,GAR009:064304}. As seen, the agreement 
between VPT2 and exact variational 
calculations is excellent, with differences being 
less than $3\,\mathrm{cm^{-1}}$. 
This is not surprising 
given the absence of large-amplitude 
fundamental modes 
and the existence of 
moderate 
mode-mode couplings (Table~\ref{tab:anharm}),    
features that clearly justify a VPT2 treatment~\citep{NIE51:90,MIL72:115,ALI85:1}.  
Yet, as noted elsewhere~\citep{FOR013:1}, discrepancies between VPT2 and VAR  
may quickly appear for overtones and 
combination bands as their energy levels and associated wavefunctions naturally sample larger sections of the molecular PESs that may not be properly described by internal-coordinate QFFs and perturbation theory. In this context, 
the use of variational approaches in conjunction with Morse-cosine (-sine) PESs/QFFs   
becomes even more critical~\citep{FOR013:1}.  

As Tables~\ref{tab:viblin}~and~\ref{tab:vibcyc} show, the computed  
VPT2 anharmonic zero-point energies (ZPEs) for $\ell$-CCN$^{-}$($^{3}\Sigma^{-}$) and $c$-CNC$^{-}$($^{1}A_{1}$) are $1851.0$~and~$1777.5\,\mathrm{cm^{-1}}$, respectively. 
With these values and considering the electronic energies we obtain 
at the corresponding QFF minima ($-130.914535769027\,\mathrm{E_{h}}$~and~$-130.889856789944\,\mathrm{E_{h}}$ for the linear and cyclic forms), an accurate estimate of   
their 0\,K energy difference can then be cast, this being~$15.3\,\mathrm{kcal\,mol^{-1}}$ as previouly indicated (note that the use of 
the associated VAR ZPEs 
has little effect on this final value). 
As expected, isotopic substitution for the heavier $^{15}$N or $^{13}$C atoms leads to significant reductions in the isotopologues' ZPE content~\citep{ROC021:A142}. This is particularly  true for $\ell$-C$^{13}$CN$^{-}$ whose ZPE 
decreases by $\approx\!34\,\mathrm{cm^{-1}}$ upon $^{13}$C replacement; similar isotopic shifts follow for other vibrational levels.
\begin{table}[htb!]
\begin{small}
\centering
\caption{\footnotesize Selected low-$J$ (hyperfine) transitions of~$\ell$-CCN$^{-}$($^{3}\Sigma^{-}$)~and~$c$-CNC$^{-}$($^{1}A_{1}$) in their ground vibrational states {\color{black}and their expected uncertainties;} see the marked lines in Figure~\ref{fig:spec}. The corresponding instruments capable of detecting them are also surveyed.}
\label{tab:lines}                                                                                                                                                                                                                                                                                                                  
\begin{threeparttable}
\begin{tabular}{
c
S[table-align-text-post=false,table-format=6.0]
S[table-align-text-post=false,table-format=3.0]@{}@{}
c}
\hline\hline \\[-1.8ex]
\multicolumn{4}{c}{$\ell$-CCN$^{-}$($^{3}\Sigma^{-}$)} \\
\hline \\[-2.25ex]
  Transition\tnote{a}                                         & {Frequency} & {Uncertainty\tnote{b}} & \multirow[b]{2}{*}{Instrument\tnote{c}} \\[0.5ex]
  $N'_{J'}(F')\leftarrow N''_{J''}(F'')$                      & {(MHz)}     & {(MHz)}                &                            \\[0.5ex]
\hline \\[-2.25ex]  
  $1_{2}(3)\leftarrow 0_{1}(2)$                            &    22869       &    15             & GBT                        \\
  $2_{3}(4)\leftarrow 1_{2}(3)$                            &    47082       &    29             & ALMA band 1/GBT            \\
  $3_{4}(5)\leftarrow 2_{3}(4)$                            &    70961       &    44             & GBT                        \\
  $4_{5}(6)\leftarrow 3_{4}(5)$                            &    94754       &    58             & ALMA band 3/GBT            \\
  $5_{6}(7)\leftarrow 4_{5}(6)$                            &   118514       &    73             &                            \\
\\
\hline \\[-2.25ex]
\multicolumn{4}{c}{$c$-CNC$^{-}$($^{1}A_{1}$)} \\
\hline \\[-2.25ex]
  Transition\tnote{a}                                         & {Frequency} & {Uncertainty\tnote{b}} & \multirow[b]{2}{*}{Instrument\tnote{c}} \\[0.5ex]
  $J'_{K_{a}'K_{c}'}(F')\leftarrow J''_{K_{a}''K_{c}''}(F'')$ & {(MHz)}     & {(MHz)}                &                            \\[0.5ex]
\hline \\[-2.25ex]  
  $2_{20}(3)\leftarrow 2_{11}(3)$                          &    26919       &    17                  & GBT                        \\
  $2_{11}(3)\leftarrow 2_{02}(3)$                          &    57621       &    36                  & GBT                        \\ 
  $1_{11}(2)\leftarrow 0_{00}(1)$                          &    64028       &    40                  & GBT                        \\
  $2_{02}(3)\leftarrow 1_{11}(2)$                          &   101135       &    63                  & ALMA band 3/GBT            \\
  $3_{13}(4)\leftarrow 2_{02}(3)$                          &   144956       &    90                  & ALMA band 4                \\
  $4_{04}(4)\leftarrow 3_{13}(3)$                          &   185944       &   115                  & ALMA band 5                \\
  $3_{31}(4)\leftarrow 2_{20}(3)$                          &   239192       &   148                  & ALMA band 6                \\
\hline\hline
\end{tabular}
\begin{tablenotes}[flushleft]
  \item[a]{{\footnotesize Only the most intense hyperfine components are reported}.}
  \item[b]{{\footnotesize Estimated uncertainties in the predicted 
  transition frequencies. The calculations assume that the theoretical $A_{0}$, $B_{0}$ and $C_{0}$ constants are accurate to within $0.062\%$ (on average) of experiment (Table~S3) so that their computed errors are $\Delta A_{0}\!\approx\!\epsilon A_{0}$, $\Delta B_{0}\!\approx\!\epsilon B_{0}$ and $\Delta C_{0}\!\approx\!\epsilon C_{0}$ with $\epsilon\!=\!6.2\times10^{-4}$. For $\ell$-CCN$^{-}$, the calculated uncertainties in the transition frequencies are $\approx\!2N'\Delta B_{0}$~(see text~and~Ref.~\citenum{SCH37:342}), while for $c$-CNC$^{-}$ they are 
  estimated using $\Delta A_{0}$, $\Delta B_{0}$ and $\Delta C_{0}$ and the energy formulae given in Table 7.7 (page~245) of Ref.~\citenum{GOR84:MMS}}.}
  \item[c]{{\footnotesize ALMA band 1 is still under construction}.}
\end{tablenotes}
\end{threeparttable}
\end{small}
\end{table}

According to Table~\ref{tab:viblin}, our best (variational) results for the 
$\ell$-CCN$^{-}$($^{3}\Sigma^{-}$) fundamentals are $1696.5\,(\nu_{1})$, $452.1\,(\nu_{2})$, 
and $1045.8\,\mathrm{cm^{-1}}\,(\nu_{3})$. Most evidently, 
the calculated VPT2 and VAR $\nu_{2}$ ($\mathrm{C\!-\!C\!-\!N}$ bend) frequencies   
are shown to match nearly perfectly 
the corresponding experimental estimate of~\citep{GAR009:064304,MUZ015:105} $452.9\pm2.9\,\mathrm{cm^{-1}}$, exhibiting errors of only $\lesssim\!1\,\mathrm{cm^{-1}}$; see Table~\ref{tab:viblin}. This is undoubtedly an asset of the present composite \emph{ab initio} energy scheme.  
Note that, in deriving the above experimental value for $\nu_{2}$ in $\ell$-CCN$^{-}$, the photoelectron spectroscopic data of Garand~\emph{et al.}~\citep{GAR009:064304} were used (see peaks~\emph{A}~and~\emph{a} therein) in combination with the 
revisited $\ell$-CCN$(^{2}\Pi)$ $(000)\,\Pi_{1/2}$\,--\,$(010)\,\mu\Sigma$ energy splitting 
reported by Muzangwa~and~Reid~\citep{MUZ015:105}. As for $c$-CNC$^{-}$($^{1}A_{1}$), the vibrational band 
origins here computed with VPT2 and our composite force field are in 
reasonable agreement with those reported from the CcCR QFF~\citep{FOR017:22860}; 
see Table~\ref{tab:vibcyc}. The largest discrepancy (of up to $\sim\!4\,\mathrm{cm^{-1}}$) 
is found for the $\mathrm{C\!-\!N}$ asymmetric stretch ($\nu_{3}$), 
a trend that becomes already clear at the harmonic level ($\omega_{3}$), {hence}  
being probably 
better explained 
by the observed variance of the corresponding diagonal 
quadratic force constant $F_{33}$ (Table~\ref{tab:force}). As noted previously, 
such disparities are undoubtedly attributed to the HO corrections [Eq.~(\ref{eq:ho})]. 
Our best estimates place the fundamental band origins of $c$-CNC$^{-}$($^{1}A_{1}$) 
at $1460.3\,(\nu_{1})$, $993.6\,(\nu_{2})$, 
and $1046.4\,\mathrm{cm^{-1}}\,(\nu_{3})$. 
It is thus {\color{black}hoped} that the results here presented aid in future high-resolution 
laboratory experiments and hopefully astronomical 
observations of $\mathrm{C_{2}N}^{-}$ as briefly surveyed next.  

\section{Astrophysical implications}\label{sec:astro}
Figure~\ref{fig:spec} shows the simulated rotational spectra of $\ell$-CCN$^{-}$($^{3}\Sigma^{-}$) and $c$-CNC$^{-}$($^{1}A_{1}$) 
at 
10\,K using~\texttt{PGOPHER}~\citep{WES017:221} and the 
spectroscopic constants presented in Tables~\ref{tab:rotlin},~\ref{tab:rotcyc},~and~S5; {\color{black}the parameters utilized here are $A_{0}$, $B_{0}$ and $C_{0}$ for the rotational constants, while the 
equilibrium values of the centrifugal distortion, fine and hyperfine structure are employed throughout, 
{hence} neglecting  
their reportedly very small 
vibrational ZPE effects~\citep{ABU006:206,CAZ016:A126,ZHA014:L28}.
} Such a low rotational 
excitation temperature is typical of those 
found in cold dense cloud cores like TMC-1; the corresponding synthetic spectra obtained at higher $T$'s characteristic of outer circumstellar envelopes of IRC+10216, $\approx\!100\,\mathrm{K}$, are depicted in Figure~S2. Also displayed for comparison are the associated working ranges of GBT and ALMA receiver {\color{black}bands on top}. A few selected low-$J$ intense lines {\color{black}and their expected uncertainties} are reported in Table~\ref{tab:lines}, wherein a direct link between the predicted rotational signatures and the 
instrument detection capabilities is also made; the associated \texttt{PGOPHER} 
files can be found in the ESI.     

As Figure~\ref{fig:spec}~(a) evinces, the rotational spectrum of $\ell$-CCN$^{-}$ is 
characteristic of a linear $^{3}\Sigma$ species, where each rotational level (except the one with $N\!=\!0$) is split into $F_{1}\,(J\!=\!N\!+\!1)$, $F_{2}\,(J\!=\!N)$ and 
$F_{3}\,(J\!=\!N\!-\!1)$ fine structure components by the presence of the  
electron spin-spin ($\lambda$) and spin-rotation ($\gamma$) interactions~\citep{SCH37:342}. Here, $J$ and $N$ are total angular momentum quantum numbers including and excluding electron spin, respectively; accurate energy formulae for such $F$ spin-triplets in terms of $J$,~$N$,~$\lambda$~and~$\gamma$ were given by Schlapp~\citep{SCH37:342}. As noted elsewhere~\citep{OHI91:L39} and clearly perceived here, rotational transitions among $F_{1}$ components show greater line strengths than those 
within the $F_{2}$ or $F_{3}$ ladders; see marked lines in Figure~\ref{fig:spec}~(a) and Table~\ref{tab:lines}. 
Inclusion of hyperfine interactions via electron spin-nuclear spin ($b_{F}$, $c$), $^{14}$N quadrupole ($eQq$), and nuclear spin-rotation ($c_{I}$) cause additional intricate splittings in the observed spectrum as the inset of Figure~\ref{fig:spec}~(a) portrays. {\color{black}These characteristic spectral signatures arising from its intrinsic (hyper)fine structure may indeed offer an extra diagnostic tool to identify this species, 
notably in radioastronomical line surveys conducted at conceivably congested frequency domains like in the 
centimeter/millimeter-wave region (3-$300\,\mathrm{GHz}$).} Note that, for simplicity, in Table~\ref{tab:lines}, only the most intense hyperfine sub-component $F_{1}$ transitions are reported; the corresponding lower and upper states are identified by their total angular momentum quantum numbers including nuclear spin, $F''$ and $F'$. To further assess the reliability of our theoretical predictions 
for $\ell$-CCN$^{-}$($^{3}\Sigma^{-}$), we again resort to  
the isoelectronic $\ell$-$\mathrm{CCO}(^{3}\Sigma^{-})$ species. 
Using its spectroscopic constants calculated by the methods described here (sections~\ref{sec:bench}~and~\ref{sec:results}), the corresponding
synthetic rotational spectrum at 10\,K has been so generated and compared with the simulated experimental one; see Figure~S3. The results have shown that our approach is capable of reproducing the well-known interstellar $\ell$-$\mathrm{CCO}$ lines $1_{2}\!\leftarrow\!0_{1}$ (22258.2\,MHz) and $2_{3}\!\leftarrow \!1_{2}$ (45826.7\,MHz) detected in TMC-1~\citep{OHI91:L39} to within 17 and 33\,MHz of experiment, respectively, {hence} posing reliable constraints on the expected 
errors for C$_{2}$N$^{-}$; see Table~\ref{tab:lines}. 
{\color{black}Such an accuracy should be sufficient to initiate astronomical line surveys on this nitrile anion, even when within the uncertainty range other and possibly unassigned transitions are found.} Of course, because the uncertainties in the theoretically predicted 
line frequencies scale roughly as $\sim\!2N'\Delta B_{0}$ [$\Delta B_{0}$ is the error in the computed rotational constant (Table~\ref{tab:lines})], one would expect to find the least deviations in the low-$N$ (low-frequency) range of 
the spectrum, \emph{i.e.}, in the centimeter and lower end of the millimeter-wave regions ($\lesssim\!150\,\mathrm{GHz}$). 
{\color{black} Note, however, that, despite influencing the predicted (low-resolution) peak positions, such 
uncertainties 
are expected to have little effect on the overall 
hyperfine splitting patterns reported herein. 
} 

As for the $c$-CNC$^{-}$ asymmetric top, the predicted spectral distribution at 10\,K 
and its most intense lines all fall within the centimeter/millimeter-wave range, 
{hence} being likewise amenable to radio observations; see Figure~\ref{fig:spec}~(b).  
Owing to the intrinsic nature of its dipole moment ($\mu$ lies in the $b$ principal axis which in turn coincides with $C_{2}$), the pure rotational spectrum of $c$-CNC$^{-}$ 
is characterized by $b$-type transitions for which $\Delta K_{a}\!=\!\pm 1$ and $\Delta K_{c}\!=\!\pm 1$~--~the quantum numbers $K_{a}$ and $K_{c}$ refer to the projection of $J$ along the molecules' figure axis in the prolate and oblate limits, respectively. Moreover, because the two equivalent off-axis C atoms are bosons, only half of the rotational levels exist, those with $K_{a}\!+\!K_{c}$ even. 
As Figure~\ref{fig:spec}~(b) and Table~\ref{tab:lines} show, {the most intense lines of such} occur in the $Q$-~and~$R$-branches~\citep{COO012:698392}. 
Inclusion of hyperfine $^{14}$N quadrupole and spin-rotation couplings have the expected effects on the observed spectrum, with the predicted splitting pattern for the 
lowest frequency intense transition $2_{20}\!\leftarrow \!2_{11}$ being shown in the inset of Figure~\ref{fig:spec}~(b). {\color{black}Again, such intrinsic (hyper)fine structure  
undoubtedly convey an additional identity to the underlying species that might be extremely useful 
to circumvent spectral line confusion, {hence} enabling 
its unambiguous identification in space.
}  

Apart from the likely detectability of $\ell$-CCN$^{-}$ and $c$-CNC$^{-}$ in the radio band with GBT, ALMA, 
and, possibly, 4GREAT (onboard the Stratospheric Observatory for Infrared Astronomy, SOFIA) at higher excitation $T$'s (Figure~S2), astronomical searches  
in the mid-/long-infrared (IR) 
should also reveal, if abundant, these molecules' rovibrational signatures.  
Indeed, their predicted band origins (Tables~\ref{tab:viblin}~and~\ref{tab:vibcyc}) are within the instrument ranges of
the Echelon-Cross-Echelle Spectrograph (EXES) onboard the SOFIA ($\approx\!2200$--$350\,\mathrm{cm^{-1}}$) as well as MIRI, the Mid-InfraRed Instrument for the James Webb Space Telescope (JWST) to be launched later this year. The present study and the highly accurate theoretical data supplied herein will certainly assist in such spectral surveys, thus paving the way for their unequivocal identification both in the laboratory and in space. {\color{black}

Apart from the main isotopologues, the data presented herein for the 
singly-substituted $^{13}$C- or $^{15}$N-bearing variants and 
their 
intrinsic (mass-shifted) spectral signatures (including hyperfine structure) add 
considerably to our C$_{2}$N$^{-}$ 
observational tool kit. These species could significantly contribute to the spectral richness of line surveys and their eventual detection would undoubtedly 
convey additional 
information on C$_{2}$N$^{-}$ 
formation pathways and chemical fractionation effects~\citep{ROC021:A142}, as well as help in gaining extra knowledge on the physical conditions (densities, temperatures and timescales) characteristic of the environment in which they form; this is the case for the recently identified rare  
isotopologues of complex organic molecules~\citep{WOON}. 
} 

\section{Conclusions}\label{sec:conclusions}
Observations of large, highly-dipolar carbon chain anions 
in a variety of interstellar environments have helped in 
establishing 
the grounds on which our current knowledge of 
ISM anion chemistry is based. 
The question remains as to whether the smallest congeners indeed play a role and, if so, how they are formed~\citep{PET96:137,COR009:68,GIA017:42,CHA020:90,YUR020:5098}. 
Studying the astronomical abundance of even smaller anions,~\emph{e.g.}, $\mathrm{C_{2}^{-}}$, $\mathrm{CH^{-}}$, $\mathrm{C_{2}H^{-}}$, $\mathrm{CN^{-}}$, and $\mathrm{C_{2}N^{-}}$, for which REA
to their parent neutrals appears to be an unlikely formation pathway,  
should then help in providing the answers. Advancing this research undoubtedly requires 
accurate knowledge of their spectroscopic signatures, 
which for the case of $\mathrm{C_{2}N^{-}}$ are as yet largely absent.  
Prompted by such a pursuit and by the recent experimental findings by~\citet{CHA020:90}, 
in this work, we 
provide such 
data 
for both ground $\ell$-CCN$^{-}$($^{3}\Sigma^{-}$) and low-lying 
$c$-CNC$^{-}$($^{1}A_{1}$) forms using 
state-of-the-art rovibrational quantum chemical techniques. Special efforts are put into the computation of
their QFFs 
by means of a high-level CC-based composite energy scheme 
that includes extrapolations to both (all-electron) one-particle and (approximate) $\mathcal{N}$-particle basis set limits, 
in addition to relativistic effects. The final analytic QFFs were 
then obtained in the usual fashion by least-squares fit, affording composite equilibrium geometries and final force constants with unprecedented accuracy. With these PESs, nuclear motion calculations have then been carried out using both 
perturbation theory and exact variational methods. Besides standard rovibrational spectroscopic constants and anharmonic vibrational 
frequencies, 
the computed data set of properties includes fine and hyperfine 
interaction constants 
evaluated \emph{ab initio} at the QFF 
equilibrium geometries and is expected to embrace 
the most 
reliable theoretical estimates to date for $\mathrm{C_{2}N^{-}}$; similar attributes 
are also provided for the $^{13}$C and $^{15}$N singly-substituted isotopologues.
The spectrocopic 
parameters so found can be readily introduced as guesses in standard experimental data reduction analyses through effective Hamiltonians. 
On the basis of benchmark calculations 
performed anew for a minimal test set of prototypical triatomics,  
%
%
%
the present protocol is shown to 
rival 
well-established methodologies 
currently available in the literature, 
producing rotational constants 
and vibrational fundamentals to within $\sim\!0.1\%$ and $\sim\!0.3\%$ of experiment, respectively, for species with at least two heavy atoms. 
This preliminary assessment thus allows for a systematic evaluation of the expected uncertainties for C$_{2}$N$^{-}$. 
By relying on their presumably similar electronic structure, 
comparisons are particularly made with the isoelectronic $\ell$-$\mathrm{CCO}(^{3}\Sigma^{-})$ radical for which accurate gas-phase experimental data are available. 
Specifically for $\ell$-CCN$^{-}$($^{3}\Sigma^{-}$), 
the calculated $\nu_{2}$ bending frequency is shown to reproduce {its associated} experimental estimate to better than $1\,\mathrm{cm^{-1}}$. 
Such accuracies reported herein should be sufficient for astronomical line surveys on C$_{2}$N$^{-}$. Using the theoretically-{predicted}  
spectroscopic constants, the rotational spectra of both $\ell$-CCN$^{-}$($^{3}\Sigma^{-}$) and $c$-CNC$^{-}$($^{1}A_{1}$) 
are derived and {the predicted transitions frequencies are  
further compared}
with working frequency ranges of powerful astronomical facilities such as GBT and ALMA. Our best theoretical estimate places $c$-CNC$^{-}$($^{1}A_{1}$) at about $15.3\,\mathrm{kcal\,mol^{-1}}$ above 
$\ell$-CCN$^{-}$($^{3}\Sigma^{-}$) 
which might limit its (astro-)chemical synthesis 
to higher-temperature environments such as in circumstellar envelopes of evolved stars and in the atmosphere of Titan. 

\section*{Conflicts of interest}
There are no conflicts to declare.

\section*{Acknowledgements}
This work has received funding from the {European Union's Horizon 2020 research and innovation program under the Marie Sklodowska-Curie} grant agreement No 894321. C.M.R.R thanks also the Academic Leiden Interdisciplinary Cluster Environment (ALICE) provided by Leiden University for the computational resources.  



\balance



\providecommand*{\mcitethebibliography}{\thebibliography}
\csname @ifundefined\endcsname{endmcitethebibliography}
{\let\endmcitethebibliography\endthebibliography}{}

\end{document}